\documentclass{JHEP3}

\usepackage{amsmath}
\usepackage{latexsym}
\usepackage{graphicx}
\usepackage{cite}
\usepackage{bbm}
\usepackage{bm}
\usepackage{mathrsfs}

\allowdisplaybreaks[1]

\numberwithin{equation}{section}

\def\AdSs5{$AdS_5$}
\def\AdSS5{$AdS_5$}
\def\AdS5s5{$AdS_5 \times S^5$}
\def\al{{\alpha^{\prime}}}
\def\gs{g_{st}}
\def\gy{g_{_{\rm YM}}}
\def\er{{\rm e}}
\def\dr{{\rm d}}
\def\Tr{{\rm Tr}}

\def\gs{g_{\rm s}}
\newcommand{\NSNS}{${\rm NS}\!\otimes\!{\rm NS}$\ }
\newcommand{\RR}{${\rm R}\!\otimes\!{\rm R}$\ }
\newcommand{\NSNSm}{{\rm NS}\otimes{\rm NS}}
\newcommand{\RRm}{{\rm R}\otimes{\rm R}}
\newcommand{\eg}{{\it e.g.~}}
\newcommand{\ie}{{\it i.e.~}}

\newcommand{\s}{\sigma}

\newcommand{\be}{\begin{equation}}
\newcommand{\ee}{\end{equation}}
\newcommand{\ba}{\begin{eqnarray}}
\newcommand{\ea}{\end{eqnarray}}
\newcommand{\bdm}{\begin{displaymath}}
\newcommand{\edm}{\end{displaymath}}
\newcommand{\ra}{\rangle}
\newcommand{\la}{\langle}
\newcommand{\pp}{\prime}

\newcommand\fr[1]{\frac{1}{#1}}
\newcommand{\barZ}{{\bar Z}}

\newcommand{\wtil}{\widetilde}
\newcommand{\what}{\widehat}
\newbox\SlashedBox
\def\fs#1{\setbox\SlashedBox=\hbox{#1}
\hbox to
0pt{\hbox to 1\wd\SlashedBox{\hfil/\hfil}\hss}{#1}}
\def\hboxtosizeof#1#2{\setbox\SlashedBox=\hbox{#1}
\hbox to
1\wd\SlashedBox{#2}}

\def\ms#1{\setbox\SlashedBox=\hbox{$#1$}
\hbox to 0pt{\hbox to
1\wd\SlashedBox{\hfil/\hfil}\hss}#1}

\def\t2{\tau_2}
\def\IZ{\relax\ifmmode\mathchoice {\hbox{\cmss Z\kern-.4em Z}}
{\hbox{\cmss Z\kern-.4em Z}}
{\lower.9pt\hbox{\cmsss Z\kern-.4em Z}}
{\lower1.2pt\hbox{\cmsss Z\kern-.4em Z}}
\else{\cmss Z\kern-.4em Z}\fi}

\def\S{\Sigma}

\def\b{\beta}
\def\a{{\alpha}}
\def\g{\gamma}
\def\veps{\varepsilon}

\def\adot{{\dot\alpha}}
\def\bdot{{\dot\beta}}

\def\d{\delta}

\def\D{\Delta}

\def\c1{{\chi^1}}

\def\v{\varphi}

\def\N4{{\cal N}=4}
\def\half{\frac{1}{2}}

\def\nn{\nonumber}

\def\nsix{(\bar\nu \nu)_{\bf 6}}

\newcommand{\tS}{\tilde{S}}
\newcommand{\bedot}{{\dot b}}
\newcommand{\calD}{{\mathcal D}}

\newcommand{\calJ}{{\mathcal J}}

\newcommand{\calZ}{{\mathcal Z}}
\newcommand{\scrA}{{\mathscr A}}

\newcommand{\scrL}{{\mathscr L}}

\newcommand{\scrN}{{\mathscr N}}
\newcommand{\scrO}{{\mathscr O}}

\DeclareMathAlphabet{\mathpzc}{OT1}{pzc}{m}{it}

\newcommand{\scrm}{{\mathpzc m}}

\newcommand{\scrmf}{{\mathpzc m}_{\rm \, f}}
\newcommand\hsp[1]{\hspace*{#1 cm}}

\newcommand{\mb}[1]{\mathbf{#1}}

\newcommand{\bfa}{\bm{a}}

\newcommand{\bfel}{\bm{\ell}}

\newcommand{\bfm}{\bm{m}}
\newcommand{\bfn}{\bm{n}}
\newcommand{\bfp}{\bm{p}}
\newcommand{\bfq}{\bm{q}}
\newcommand{\bfr}{\bm{r}}
\newcommand{\bfs}{\bm{s}}
\newcommand{\bfu}{\bm{u}}
\newcommand{\bfv}{\bm{v}}

\newcommand{\bfi}{\bm{i}}
\newcommand{\bfj}{\bm{j}}
\newcommand{\bfal}{\bm{\alpha}}

\title{Mixing of the RR and NSNS sectors in the BMN limit}
\author{Michael B. Green and Aninda Sinha \\
Department of Applied Mathematics and
Theoretical Physics \\
Wilberforce Road, Cambridge CB3 0WA, UK \\
E-mail: \email{M.B.Green@damtp.cam.ac.uk, A.Sinha@damtp.cam.ac.uk}}
\author{Stefano Kovacs \\
Max-Planck-Institut f\"ur Gravitationsphysik \\
Albert-Einstein-Institut \\
Am M\"uhlenberg 1, D-14476 Potsdam, Germany  \\
E-mail: \email{stefano.kovacs@aei.mpg.de}}

\abstract{This paper concerns instanton contributions to two-point
correlation functions of BMN operators in $\scrN$=4 supersymmetric
Yang--Mills that vanish in planar perturbation theory. Two-point
functions of operators with even numbers of fermionic impurities (dual
to \RR string states) and with purely scalar impurities (dual to \NSNS
string states) are considered. This includes mixed \RR\!\!--\NSNS
two-point functions. The gauge theory correlation functions are shown
to respect BMN scaling and their behaviour is found to be in good
agreement with the corresponding $D$-instanton contributions to
two-point amplitudes in the maximally supersymmetric IIB plane-wave
string theory. The string theory calculation also shows a simple
dependence of the mass matrix elements on the mode numbers of states
with an arbitrary number of impurities, which is difficult to extract
from the gauge theory. For completeness, a discussion is also given of
the perturbative mixing of two-impurity states in the \RR and \NSNS
sectors at the first non-planar level.}

\preprint{DAMTP-2005-101 \\
AEI-2005-151 \\
{\tt hep-th/0512198}}

\begin{document}

\section{Introduction and summary}
\label{intro}

The correspondence between string theory in a maximally supersymmetric
IIB plane-wave background \cite{penrose} and the BMN sector of the
$\scrN$=4 supersymmetric Yang--Mills (SYM) theory \cite{bmn} has been
extensively studied at the perturbative level. Non-perturbative
aspects of the duality have recently been analysed in \cite{gks1} and
\cite{gks2}, where it was shown that the striking agreement between
the effects of $D$-instantons and of Yang--Mills instantons, found in
the original formulations of the AdS/CFT correspondence
\cite{bgkr,dhkmv,gk}, persists in the BMN/plane-wave limit.  This
paper extends this analysis to include bosonic states with an even
number of fermionic impurities in the gauge theory and the
corresponding \RR states in the dual string theory.  The further
extension to include fermionic states (which have an odd number of
fermionic impurities)  involves a straightforward generalisation of
these results.

In the BMN limit the gauge theory -- string theory correspondence
relates the string mass spectrum to the spectrum of scaling dimensions
of Yang--Mills gauge invariant operators of large dimension, $\D$, and
large charge, $J$, with respect to a U(1) subgroup of the SU(4)
R-symmetry group. This relation is formally realised via the operator
identity
\be
\fr{\mu} \, H^{(2)} = \calD-\calJ \, ,
\label{h-dil-id}
\ee
relating the string theory hamiltonian to the combination
$\calD-\calJ$ of the gauge theory dilation operator and U(1)
generator. The duality involves the double limit, $\D\to\infty$,
$J\to\infty$, with $\D-J$ kept finite, on the eigenvalues of the
operators $\calD$ and $\calJ$. The parameter $\mu$ in (\ref{h-dil-id})
is related to the mass parameter, $m$, entering the light-cone string
action by $m=\mu\al p_-$ (where $p_-$ is the light-cone momentum) and
equals the background value of the \RR five-form. The equality
(\ref{h-dil-id}) implies that the eigenvalues of the operators on the
two sides should coincide. Numerous tests of this relation have been
carried out at the perturbative level
\cite{pert,recentgauge,kpss,mit1,bkpss,mit2,sz}. In \cite{gks1,gks2}
$D$-instanton contributions to the plane-wave string mass matrix for
certain states with up to four bosonic string excitations were shown
to be in striking agreement with instanton contributions to the matrix
of anomalous dimensions in the corresponding sectors of the dual gauge
theory. A brief review of these results is presented in \cite{gks3}.

In the large $N$ limit and in the BMN sector of the gauge theory
the r\^ole of the ordinary 't Hooft parameters, $\lambda$ and $1/N$,
is played by effective rescaled parameters \cite{kpss,mit1},
\be
\lambda^\pp = \frac{\gy^2N}{J^2}  \, , \qquad g_2 = \frac{J^2}{N} \, .
\label{eff-param}
\ee
In the BMN correspondence these are related to the string parameters
via
\be
m^2 = (\mu p_-\al)^2 = \fr{\lambda^\pp} \, ,
\qquad 4\pi\gs m^2 = g_2 \, ,
\label{param-ids}
\ee
which imply that in the double scaling limit, $N\to\infty$,
$J\to\infty$, with $J^2/N$ fixed, the weak coupling regime of the
gauge theory corresponds to the limit of small $g_s$ and large $m$ on
the string side.

The string hamiltonian is the sum of two pieces,
\be
H^{(2)} = H^{(2)}_{\rm pert} + H^{(2)}_{\rm non-pert} \, .
\label{pert+npert-h}
\ee
The perturbative part has an expansion in powers of $g_s$, which gets
reorganised into a series in $g_2$. The non-perturbative part contains
the $D$-instanton induced corrections. In the BMN limit of the
$\scrN$=4 SYM theory, after the operator mixing is resolved
\cite{bers}, quantum corrections to the eigenvalues of $\calD-\calJ$
are also expected to be organised in a double series in $\lambda^\pp$
and $g_2$ (a property referred to as BMN scaling), with $g_2$ playing
the r\^ole of genus counting parameter. According to \cite{bmn} the
$g_2$ expansion in string theory is term by term exact to all orders
in $\lambda^\pp$. This means that the free string spectrum is
identified with the resummed planar expansion of the spectrum of the
$\calD-\calJ$ operator  on the gauge side. Loop corrections in string
theory correspond to non-planar effects in the Yang--Mills theory. At
each order in the loop expansion, the string theory encodes an
infinite series of $\lambda^\pp$ corrections in the gauge theory at
the fixed corresponding order in $g_2$.

The large body of work on perturbative and non-perturbative
contributions to anomalous dimensions of BMN operators has
concentrated almost entirely on states with bosonic impurities.
Correspondingly, almost all results on the plane-wave string mass
spectrum refer to strings with bosonic excitations.  However,
fermionic impurities are obviously required in  any complete treatment
of the mass matrix.  States with an even number of fermionic
impurities correspond to \RR states of the string theory.  In general
one would expect  such states to mix with those containing bosonic
impurities, or \NSNS states in the string description.  Indeed, in
\cite{gks1} it was noted that certain string two-point functions that
mix the \NSNS and \RR sectors receive non-zero $D$-instanton
contributions even though these states do not mix at tree level. In
this paper we will study these classes of string amplitudes in detail,
together with  the dual correlation functions in the BMN limit of
$\scrN$=4 SYM. The string states and gauge theory operators that we
consider contain an arbitrary even number of fermionic and bosonic
impurities, but in specific combinations.

On the gauge theory side we find that the two-point functions respect
BMN scaling and we determine their dependence on the parameters,
$\lambda^\pp$ and $g_2$, in the semi-classical approximation.
Interestingly, we find that, depending on the number and combination
of impurities, the result can contain arbitrarily large inverse powers
of $\lambda^\pp$. The dual string amplitudes, computed using the
formalism of \cite{gks2}, are shown to be in very good agreement with
the gauge theory results. The string theory calculation also shows a
remarkably simple dependence of the mass matrix elements on the mode
numbers of states with an arbitrary number of impurities. The
dependence on the mode numbers is extremely complicated to determine
through a standard instanton calculation in the Yang--Mills theory and
thus the string result represents a highly non-trivial prediction for
the gauge side.

The mixing of the \NSNS and \RR sectors can easily be motivated from
the presence of background \RR flux in the string picture. First note
that \RR charge conservation is violated in tree-level closed-string
scattering from a $D3$-brane, so that \NSNS and \RR states mix at tree
level in AdS$_5\times S^5$ (which is the near-horizon geometry of a
stack of coincident $D3$-branes). The Penrose boost that takes
AdS$_5\times S^5$ to the maximally supersymmetric IIB plane-wave
background leads to a string theory in which \RR charge is conserved
on a spherical world-sheet (tree level). However, the non-zero
background flux (non-zero $\mu$) leads to the possibility of mixing
\NSNS and \RR states by string loop corrections, as will be indicated
later in this paper. This should mean that non-planar perturbative
contributions in the gauge theory (\ie beyond the zeroth order in the
$g_2$ expansion) mix states that have bosonic impurities with states
that have an even number of fermionic impurities. We will later show
that this is indeed the case by analysing the leading planar and
non-planar contributions to a specific mixed two-point function.

The paper is organised as follows. In section \ref{bmnops} we define
the different classes of BMN operators which we focus on and we
explain our notation. Section \ref{2ptfuncts} discusses instanton
contributions to Yang--Mills two-point functions in the semi-classical
approximation. The calculation of the dual $D$-instanton induced
amplitudes in string theory is presented in section
\ref{string2pt}. Section \ref{perturb} discusses the issue of the
perturbative mixing of the \NSNS and \RR sectors through a qualitative
analysis of a specific process.

\section{BMN operators}
\label{bmnops}

In this section we discuss certain classes of BMN operators whose
two-point functions we shall analyse in the following sections. We
consider bosonic operators which are SO(4)$_C\times$SO(4)$_R$
singlets, corresponding both to \RR states, \ie with an even number of
fermionic impurities, and to \NSNS states, \ie containing only bosonic
impurities. The operators we consider contain an arbitrary number of
impurities, but in certain specific combinations. As will be discussed
in the following, in the case of \RR states it is convenient to study
operators which also contain four bosonic impurities. The inclusion of
the bosonic impurities simplifies the analysis in the one-instanton
sector because they allow to soak up the fermion superconformal modes
without the need to use higher order solutions for any of the fields.

The operators we focus on involve scalar or fermion impurities in
singlet combinations. In the BMN limit the four scalars, $\v^i$, not
charged under U(1) transform in the
$\left[(0,0);\left(\half,\half\right)\right]$ of
SO(4)$_C\times$SO(4)$_R$. The $\scrN$=4 fermions, $\lambda^A_\a$ and
$\bar\lambda_A^\adot$, transforming in the $\mb4$ and $\mb{\bar 4}$ of
SU(4), are decomposed as \cite{gks2}
\ba
&& \lambda^A_\a \to \psi^{-\,a}_\a \oplus \bar\psi^+_{\a a} \, ,
\hsp{1} a=1,4 \label{lambdadec} \\
&& \bar\lambda^\adot_A \to \psi_{\dot a}^{+\,\adot} \oplus
\bar\psi^{-\,\adot\dot a} \, , \hsp{0.7} {\dot a}=2,3 \, ,
\label{blmabdadec}
\ea
where $\psi^{-\,a}_\a$ and $\psi^{+\,\adot}_{\dot a}$ have U(1) charge
$+\half$, \ie $\D-J=1$, whereas $\bar\psi^+_{\a a}$ and
$\bar\psi^{-\,\adot\dot a}$ have charge $-\half$, \ie $\D-J=2$. Under
the SO(4)$_C\times$SO(4)$_R$ symmetry the fermions $\psi^{-\,a}_\a$
and $\bar\psi^+_{\a a}$ transform in the
$\left[\left(\half,0\right);\left(\half,0\right)\right]$, while
$\psi_{\dot a}^{+\,\adot}$ and $\bar\psi^{-\,\adot\dot a}$ transform
in the $\left[\left(0,\half\right);\left(0,\half\right)\right]$.  The
definition of the fermions $\bar\psi^+_{\a a}$ and
$\bar\psi^{-\,\adot\dot a}$ involves the multiplication by a matrix
which flips the SO(4)$_R$ chirality and respectively lowers or raises
the corresponding index. The decomposition in
(\ref{lambdadec})-(\ref{blmabdadec}) corresponds to the decomposition
of the left- and right-moving type IIB fermions into chiral
SO(4)$_C\times$SO(4)$_R$ fermions \cite{met,mt}, $(S^-,S^+)$ and
$(\tilde S^-,\tilde S^+$).

Fermion impurities in BMN operators are associated with the insertion
of the $\D-J=1$ fields, $\psi^{-\,a}_\a$ and $\psi^{+\,\adot}_{\dot
a}$. In the dual string theory this corresponds to the insertion of
$S^-_{-n}$ and $S^+_{-n}$ (or $\tilde S^-_{-n}$ and  $\tilde
S^+_{-n}$) creation operators. The conjugate fermions,
$\bar\psi^+_{\a a}$ and $\bar\psi^{-\,\adot\dot a}$, which have
$\D-J=2$, enter into the conjugate operators. In perturbation theory
the  only non-zero contractions are between a fermion and its
conjugate, \ie $\la\psi^{-\,a}_\a\bar\psi^{-\,\bdot\dot b}\ra$ and
$\la\psi^{+\,\adot}_{\dot a} \bar\psi^+_{\b b}\ra$. This will be
important in the analysis of correlation  functions in the next
sections.

The most general BMN operator with fermionic impurities that we shall
consider is of the form
\ba
\hsp{-0.5} \scrO^{k,h}_{\bfel,\bfn,\bfm} &=&
t_{\bfi,\bfa,\bm{\bdot}}^{{\scriptscriptstyle(\RRm)}\,\bm{\bedot},\bfal}
\: c_{k,h}(\gy,N,J) \nn \\
&\!\times\!& \hspace*{-0.3cm} \begin{array}[t]{c}
{\displaystyle \sum_{p_2,p_3,p_4,u_1,\dots,v_{2h}=0}^J} \\
{\scriptstyle p_2+p_3+p_4+u_1+\cdots+v_{2h}\le J} \\
{\scriptstyle p_1=J-(p_2+p_3+p_4+u_1+\cdots+v_{2h})}
\end{array} \hspace*{-0.3cm}
e(\bfp,\bfu,\bfv;\bfel,\bfn,\bfm;J) \:
\Tr\left[\calZ^{(\bfp,\bfi)}_\v \calZ^{(\bfu,\bfal,\bfa)}_{\psi^-}
\calZ^{(\bfv,\bm\bdot,\bm{\bedot})}_{\psi^+} \right] ,
\rule{0pt}{16pt}
\label{genBMNop}
\ea
where various sets of indices have been grouped into `vectors',
\ba
&& \hsp{-1} \bfp = (p_1,p_2,p_3,p_4) \, , \hsp{1.43}
\bfu = (u_1,u_2,\ldots,u_{2k}) \, , \hsp{1.35}
\bfv = (v_1,v_2,\ldots,v_{2h}) \, , \nn \\
&& \hsp{-1} \bfel = (\ell_1,\ell_2,\ell_3) \, , \hsp{2.11}
\bfn = (n_1,n_2,\ldots,n_{2k}) \, , \hsp{1.3}
\bfm = (m_1,m_2,\ldots,m_{2h}) \, , \nn \\
&& \hsp{-1} \bfi = (i_1,i_2,i_3,i_4) \, , \hsp{1.76}
\bfa = (a_1,a_2,\ldots,a_{2k}) \, , \hsp{1.45}
\bm\bedot = (\dot b_1,\dot b_2,\ldots,\dot b_{2h}) \nn \\
&& \hsp{-1} \bm{\a} = (\a_1,\a_2,\ldots,\a_{2k}) \, ,
\hsp{0.92} \bm{\bdot} = (\bdot_1,\bdot_2,\ldots,\bdot_{2h}) \, ,
\label{rrindices}
\ea
and we have introduced the notation
\be
\calZ_\v^{(\bfp,\bfi)} = \prod_{r=1}^4 Z^{p_r}\v^{i_r} \, , \qquad
\calZ_{\psi^-}^{(\bfu,\bfal,\bfa)} = \prod_{r=1}^{2k} Z^{u_r}
\psi^{-\,a_r}_{\a_r} \, , \qquad
\calZ_{\psi^+}^{(\bfv,\bm{\bdot},\bm{\bedot})}
= \prod_{r=1}^{2h} Z^{v_r} \psi^{+\,\bdot_r}_{\dot b_r} \, .
\label{calzdef}
\ee
The coefficient $e(\bfp,\bfu,\bfv;\bfel,\bfn,\bfm;J)$ in
(\ref{genBMNop}) is given by
\ba
&& \hsp{-0.4} e(\bfp,\bfu,\bfv;\bfel,\bfn,\bfm;J)
= \exp\left\{2\pi i[p_2(\ell_1+\cdots+m_{2h})+p_3(\ell_2+\cdots+m_{2h})
\right. \nn \\
&& \hsp{1} + p_4(\ell_3+\cdots+m_{2h})+u_1(n_1+\cdots+m_{2h})
+\cdots+ u_{2k}(n_{2k}+\cdots+m_{2h}) \nn \\
&& \hsp{0.95}\left.+v_1(m_1+\cdots+m_{2h})+\cdots+v_{2h}m_{2h}]/J\right\}
\nn \, .
\label{phase}
\ea
The tensor $t_{\bfi,\bfa,\bm{\bdot}}^{{\scriptscriptstyle(\RRm)}\,
\bm{\bedot},\bfal}$ projects onto the SO(4)$_R$ singlet,
\be
t_{\bfi,\bfa,\bm{\bdot}}^{{\scriptscriptstyle(\RRm)}\,\bm{\bedot},\bfal}
= \veps_{i_1i_2i_3i_4}\,\prod_{r=1}^k\veps_{a_{2r-1}a_{2r}}
\veps^{\a_{2r-1}\a_{2r}}\, \prod_{s=1}^{h}
\veps^{\dot b_{2s-1}\dot b_{2s}} \veps_{\bdot_{2s-1}\bdot_{2s}}
\label{gentensor}
\ee
 and the normalisation coefficient,
$c_{k,h}(\gy,N,J)$, is
\be
c_{k,h}(\gy,N,J) =
\fr{\sqrt{J^{3+2k+2h}\left(\frac{\gy^2N}{8\pi^2}\right)^{J+4+2k+2h}}} \, .
\label{gennorm}
\ee
The form of the conjugate operator is similar to (\ref{genBMNop}) with
the $Z$'s replaced by $\barZ$'s and the $\psi^-$'s and $\psi^+$'s
replaced respectively by $\bar\psi^-$'s and $\bar\psi^+$s'. In the
following we shall consider two-point functions of the form
$\la\scrO^{k,h}_{\bfel,\bfn,\bfm}(x_1)\,
\bar\scrO^{k^\pp,h^\pp}_{\bfel^\pp,\bfn^\pp,\bfm^\pp}(x_2)\ra$, where
the operator $\scrO^{k,h}_{\bfel,\bfn,\bfm}$ contains $k$ $\psi^-$
and $h$ $\psi^+$ pairs and the operator
$\bar\scrO^{k^\pp,h^\pp}_{\bfel^\pp,\bfn^\pp,\bfm^\pp}$ contains
$k^\pp$ $\bar\psi^-$ and $h^\pp$ $\bar\psi^+$ pairs. The normalisation
of the operators is such that two-point functions of this type (if
non-zero) are of order 1 at tree level.

The string states which we are interested in, dual to operators of the
form (\ref{genBMNop}), are schematically, up to an overall
normalisation, of the form (see \cite{gks1} for notation)
\be
\veps_{i_1i_2i_3i_4}\,\a^{i_1}_{-\ell_1}\a^{i_2}_{-\ell_2}
\tilde\a^{i_3}_{-\ell_1}\tilde\a^{i_4}_{-\ell_2}
\left[S^-_{-n_1}\tilde S^-_{-n_1}\right]\ldots
\left[S^-_{-n_k}\tilde S^-_{-n_k}\right]\!\left[S^+_{-m_1}
\tilde S^+_{-m_1}\right]\ldots \left[S^+_{-m_h}\tilde S^+_{-m_h}
\right] |0\ra_h \, ,
\label{genstringstate}
\ee
where $|0\ra_h$ denotes the BMN ground state and the square brackets
indicate contraction of the SO(4)$_C\times$SO(4)$_R$ indices. Notice
that in (\ref{genstringstate}) we have inserted the same number of
left- and right-moving oscillators and we have chosen the mode numbers
carried by the creation operators to be equal in pairs. More general
states satisfying the physical level-matching condition can be
constructed, but we restrict our attention to those of the form
(\ref{genstringstate}) because these form a class of states that
couple to a $D$-instanton in the plane-wave background.

In the operator (\ref{genBMNop}) $k$ pairs of $\psi^-$ fermions and
$h$ pairs of $\psi^+$ fermions are contracted into
SO(4)$_C\times$SO(4)$_R$ singlets. In the operator
$\bar\scrO^{k^\pp,h^\pp}_{\bfel^\pp,\bfn^\pp,\bfm^\pp}$ the $2k^\pp$
$\bar\psi^-$'s and the $2h^\pp$ $\bar\psi^+$'s are similarly paired in
singlets. The unique singlet that can be constructed in this way
involves contractions of both types of SO(4) indices via $\veps$
tensors, see (\ref{gentensor}). This implies that the fermions are
automatically pairwise antisymmetrised in the colour indices. In the
string state (\ref{genstringstate}) there is no analogue of the colour
antisymmetrisation, but the contraction is allowed because the two
fermions in each pair are different, being a left- and a right-mover.

The other class of operators we consider are dual to string states in
the \NSNS sector. These involve an arbitrary number of scalar
impurities contracted into a SO(4)$_C\times$SO(4)$_R$ singlet. Using
the same  notation introduced in (\ref{genBMNop}) the operators are
\be
\scrO^l_{\bfel,\bfn} = t_{\bfi,\bfj}^{\scriptscriptstyle(\NSNSm)}
\: c_l(\gy,N,J+l) \hspace*{-0.5cm} \begin{array}[t]{c}
{\displaystyle \sum_{p_2,p_3,p_4,q_1,\dots,q_{2l}=0}^{J+l}} \\
{\scriptstyle p_2+p_3+p_4+q_1+\cdots+q_{2l}\le J+l} \\
{\scriptstyle p_1=J+l-(p_2+p_3+p_4+q_1+\cdots+q_{2l})}
\end{array} \hspace*{-0.5cm}
e(\bfp,\bfq;\bfel,\bfn;J+l) \;
\Tr\left[\calZ^{(\bfp,\bfi)}_\v \calZ^{(\bfq,\bfj)}_\v\right],
\rule{0pt}{16pt}
\label{genBMNopNS}
\ee
where a vector notation for the indices has been used,
\ba
&& \hsp{-1} \bfp = (p_1,p_2,p_3,p_4) \, , \hsp{1.43}
\bfq = (q_1,q_2,\ldots,q_{2l}) \, , \nn \\
&& \hsp{-1} \bfel = (\ell_1,\ell_2,\ell_3) \, , \hsp{2.11}
\bfn = (n_1,n_2,\ldots,n_{2l}) \, , \nn \\
&& \hsp{-1} \bfi = (i_1,i_2,i_3,i_4) \, , \hsp{1.76}
\bfj = (j_1,j_2,\ldots,j_{2l}) \, .
\label{nsindices}
\ea
The tensor $t_{\bfi,\bfj}^{\scriptscriptstyle(\NSNSm)}$, which
projects onto the SO(4)$_R$ singlet, is
\be
t_{\bfi,\bfj}^{\scriptscriptstyle(\NSNSm)} = \veps_{i_1i_2i_3i_4}
\, \d_{j_1j_2}\d_{j_3j_4}\cdots\d_{j_{2l-1}j_{2l}} \, ,
\label{tensorns}
\ee
\ie we choose singlet operators in which four scalars are contracted
via an $\veps$-tensor and the remaining $2l$ scalars are contracted
pairwise via Kronecker $\d$'s. The normalisation factor in
(\ref{genBMNopNS}) is
\be
c_l(\gy,N,J+l) =
\fr{\sqrt{J^{3+2l}\left(\frac{\gy^2N}{8\pi^2}\right)^{J+4+3l}}} \,
\label{gennormns}
\ee
and the phase factor in the sum, $e(\bfp,\bfq;\bfel,\bfn;J+l)$, is
\ba
e(\bfp,\bfq;\bfel,\bfn;J+l)
&=& \exp\left\{2\pi i[p_2(\ell_1+\cdots+n_{2l})+p_3(\ell_2+
\cdots+n_{2l})+p_4(\ell_3+\cdots+n_{2l}) \right. \nn \\
&+& \left. q_1(n_1+\cdots+n_{2l})+\cdots+
q_{2l}n_{2l}]/J\right\} \, .
\label{phasens}
\ea
Notice that the operator (\ref{genBMNopNS}) contains a total of $J+l$
$Z$ fields. This is necessary in order to give it the same dimension
and U(1) charge as operators with a total of $2l$ fermionic impurities
($\psi^-$ and/or $\psi^+$) of the type defined in (\ref{genBMNop}).
The number of $Z$'s in the operator is reflected in the power of
$\gy^2N/8\pi^2$ in the normalisation (\ref{gennormns}).

The string states dual to operators of the form (\ref{genBMNopNS}) are
(up to a normalisation factor)
\be
\veps_{i_1i_2i_3i_4} \, \a^{i_1}_{-\ell_1}\a^{i_2}_{-\ell_2}
\tilde\a^{i_3}_{-\ell_1}\tilde\a^{i_4}_{\ell_2}\a^{j_1}_{-n_1}
\tilde\a^{j_1}_{-n_1}\a^{j_2}_{-n_2}\tilde\a^{j_2}_{-n_2}
\cdots \a^{j_l}_{-n_l}\tilde\a^{j_l}_{-n_l} |0\ra_h \, ,
\label{genstrstatens}
\ee
where, as in the case of the \RR state (\ref{genstringstate}), we have
restricted the attention to a class of states that couple to a
$D$-instanton in the plane-wave background: the number of left- and
right-movers in (\ref{genstrstatens}) is the same and the
corresponding mode numbers are equal in pairs.

In order to construct gauge theory operators that can be identified
with the string states (\ref{genstringstate}) and
(\ref{genstrstatens}) one needs to consider linear combinations of
operators such as those defined in (\ref{genBMNop}) in
(\ref{genBMNopNS}). Since the creation operators in the string states
commute, it is necessary to sum, in the corresponding operators, over
all the possible permutations of the $\v$ and $\psi^\pm$
impurities. As in the four impurity case studied in \cite{gks2} it is
also necessary to (anti-)symmetrise the operators under permutations
of the mode numbers so that they possess the same symmetry properties
as the dual string states. This should automatically impose the
constraint that the instanton contribution vanishes unless the mode
numbers in the operator are equal in pairs as in
(\ref{genstringstate}) and (\ref{genstrstatens}). We shall not discuss
these aspects here since we shall not analyse the mode number
dependence in the gauge theory two-point functions.

The insertion of $\D-J=2$ impurities is, in general, necessary to
define well-behaved BMN operators, as observed already in the case of
two impurity operators in \cite{bkpss,bei}. This is also the case for
operators in the class we are considering. Specifically, the complete
definition of the operators should also involve terms with
$\bar\psi^+_{\a a}$ and $\bar\psi^{-\,\adot a}$ insertions as well as
terms in which pairs of $\v$'s in a singlet are replaced by a $Z\barZ$
insertion. However, these terms are not relevant at leading order in
the large $J$ and large $N$ limit and only need to be taken into
account when $g_2$ corrections are computed, \ie beyond the
semi-classical approximation.

\section{Gauge theory two-point functions in the one-instanton sector}
\label{2ptfuncts}

In this section we briefly review the calculation of one-instanton
contributions to two-point correlation functions in semi-classical
approximation and then discuss examples involving the operators
defined in the previous section.

\subsection{Semi-classical approximation}
\label{semicl}

The one-instanton contribution to the two-point correlation function
of composite operators $\scrO_1$ and $\scrO_2$ in the semi-classical
approximation takes the form
\be
G^{\rm 1-inst}(x_1,x_2) = \int \dr \mu_{\rm inst}
(\scrm_{\rm \,b},\scrm_{\rm \,f}) \, \er^{-S_{\rm inst}} \;\,
\what\scrO_1(x_1;\scrm_{\rm \,b},\scrm_{\rm \,f})
\;\what\scrO_2(x_2;\scrm_{\rm \,b},\scrm_{\rm \,f}) \, ,
\label{semiclass}
\ee
where we have denoted the bosonic and fermionic collective coordinates
by $\scrm_{\rm \,b}$ and $\scrm_{\rm \,f}$ respectively. In
(\ref{semiclass}) $\dr\mu_{\rm inst}(\scrm_{\rm \,b}, \scrm_{\rm\,f})$
is the integration measure on the instanton moduli space, $S_{\rm
inst}$ is the classical action evaluated on the instanton solution and
$\what\scrO_1$ and $\what\scrO_2$ denote the classical expressions for
the operators computed in the instanton background. In the case of
gauge-invariant operators the semi-classical expression
(\ref{semiclass}) involves the integration over the position and size
of the instanton, $x_0$ and $\rho$, and over the sixteen fermion
moduli associated with the broken supersymmetries, $\eta^A$ and
$\bar\xi^A$. The bosonic moduli associated with global gauge
orientations are integrated out. The corresponding fermion moduli,
$\nu^A$ and $\bar\nu^A$, appear in gauge-invariant operators in
colour-singlet bilinears and the integration over these moduli is
re-expressed in terms of an integration over bosonic auxiliary
variables, $\Omega^{AB}$, parametrising a five-sphere. Instanton
contributions to two-point functions of scalar operators in $\scrN$=4
SYM have been analysed in \cite{sk}. Details of the calculation of
two-point functions of BMN operators were discussed in \cite{gks2}
following the general analysis of \cite{dhkmv}. Comprehensive reviews
of instanton calculus in supersymmetric gauge theories can be found in
\cite{akmrv,dhkm,bvv}. For a generic two-point function one finds
\ba
\la\scrO_1(x_1)\,\scrO_2(x_2)\ra = \a(p,q,N) \,
\gy^{8+p+q} \, \er^{2\pi i\tau} \int \dr\rho\,\dr^4x_0\,\dr^5\Omega
\prod_{A=1}^4 \dr^2\eta^A\dr^2\bar\xi^A \, \rho^{p+q-5} && \nn \\
\times \what\scrO_1(x_1;\rho,x_0,\Omega,\eta,\bar\xi)
\;\what\scrO_2(x_2;\rho,x_0,\Omega,\eta,\bar\xi) && \!\! ,
\label{gensemicl}
\ea
where the $\bar\nu^A\nu^B$ bilinears in the operator profiles have
been re-written in terms of the auxiliary variables $\Omega^{AB}$. In
the large-$N$ limit
\be
\a(p,q,N) \sim \frac{N^{\half(p+1)}}{\pi^{p+q+\half}}
  \left[1+O(1/N)\right] \, ,
\label{largeNdep}
\ee
where $p$ and $q$ are the numbers of antisymmetric and symmetric
$\bar\nu^A\nu^B$ bilinears respectively.

In all the examples that we shall consider the classical profiles of
the operators take a factorised form. In such expressions the terms
which contribute to two-point functions can be written schematically
as
\be
\what\scrO(x;x_0,\rho,\eta,\bar\xi,\Omega) \sim f(x;x_0,\rho) \,
g(\Omega) \prod_{A=1}^4 \left[\zeta^A(x)\right]^2 \, ,
\label{factorop}
\ee
where $\zeta^A$ is a combination of the fermion modes, $\eta^A$ and
$\bar\xi^A$, associated with the broken superconformal symmetries,
\be
\zeta^A_\a(x) = \fr{\sqrt{\rho}} \left[\rho \, \eta^A_\a - (x-x_0)_\mu
\s^\mu_{\a\adot}\bar\xi^{\adot A} \right] \, .
\label{zetadef}
\ee
The generic two-point function thus becomes
\ba
\la \scrO_1(x_1)\,\scrO_2(x_2)\ra_{\rm inst} &\sim& \a(p,q,N) \,
\gy^{8+p+q} \, \er^{2\pi i\tau} \int \dr\rho\,\dr^4x_0\,
\rho^{p+q-5} \, f_1(x_1;x_0,\rho)\, f_2(x_2;x_0,\rho) \nn \\
&\times& \int \dr^8 \eta \,\dr^8\bar\xi \,\prod_{A=1}^4
\left[\zeta^A(x_1)\right]^2 \left[\zeta^A(x_2)\right]^2
\int \dr^5\Omega \, g_1(\Omega) \, g_2(\Omega) \, .
\label{2ptfactor}
\ea
After this factorisation the bosonic integration over $x_0$ and $\rho$
is logarithmically divergent and needs to be regularised. This signals
a contribution to the matrix of anomalous dimensions which is
extracted from the coefficient of the logarithmically divergent
term.

The integration over the superconformal modes in the second line of
(\ref{2ptfactor}) is straightforward,
\be
\int \dr^8 \eta \,\dr^8\bar\xi \,\prod_{A=1}^4
\left[\zeta^A(x_1)\right]^2 \left[\zeta^A(x_2)\right]^2 =
(x_1-x_2)^8 \, .
\label{suconfint}
\ee
Finally, as will be shown in the next section, the five-sphere
integrals in all the cases we are interested in can be reduced to the
form
\be
I_{S^5}(a,b,c) = \int\dr^5\Omega \, \left(\Omega^{14}\Omega^{23}
\right)^a \left(\Omega^{12}\Omega^{34}\right)^b
\left(\Omega^{13}\Omega^{24}\right)^c \, ,
\label{gen5sph}
\ee
where $a$, $b$ and $c$ are integers. This integral is a generalisation
of those encountered in the case of two and four impurity operators
and can be evaluated using the same method described in \cite{gks2}.
Defining
\ba
&& \Omega = \S^{14}_i\Omega^i = \Omega^1+i\Omega^4 \, , \qquad
\bar\Omega = \S^{23}_i\Omega^i = \Omega^1-i\Omega^4 \, , \nn \\
&& \wtil\Omega = \S^{12}_i\Omega^i = \Omega^3+i\Omega^6 \, ,
\qquad \bar{\wtil\Omega} = \S^{34}_i\Omega^i =
\Omega^3-i\Omega^6 \, , \nn \\
&& \what\Omega = \S^{13}_i\Omega^i = \Omega^2+i\Omega^5 \, ,
\qquad \bar{\what\Omega} = \S^{24}_i\Omega^i =
\Omega^2-i\Omega^5 \, ,
\label{omegasdef}
\ea
the integral (\ref{gen5sph}) can be rewritten as
\be
I_{S^5}(a,b,c) = \int \dr\Omega\dr\bar\Omega\dr\wtil\Omega
\dr\bar{\wtil\Omega}\dr\what\Omega\dr\bar{\what\Omega} \,
\d(\Omega\bar\Omega+\wtil\Omega\bar{\wtil\Omega}+
\what\Omega\bar{\what\Omega}-1) \, \left(\Omega\bar\Omega\right)^a
\left(\wtil\Omega\bar{\wtil\Omega}\right)^b
\left(\what\Omega\bar{\what\Omega}\right)^c \, .
\label{5sphabc}
\ee
This can be easily computed generalising the calculations of
\cite{gks2}. The result is
\be
I_{S^5}(a,b,c) = \pi^3 \,
\frac{\Gamma(a+1)\,\Gamma(b+1)\,\Gamma(c+1)}{\Gamma(a+b+c+3)} \, .
\label{5sphabc-res}
\ee
In the next subsections we shall apply this general analysis to
certain classes of two-point functions of the operators introduced
in section \ref{bmnops}.

\subsection{A class of two-point functions in the \RR sector}
\label{RR2pt}

In this section we analyse the one-instanton contribution to two-point
functions of the operators $\scrO^{k,h}_{\bfel,\bfn,\bfm}$ and
$\bar\scrO^{k^\pp,h^\pp}_{\bfel^\pp,\bfn^\pp,\bfm^\pp}$ defined in
section \ref{bmnops}. The generic two-point function in this class is
\be
G^{k,h;k^\pp,h^\pp}_{\bfel,\bfn,\bfm;\bfel^\pp,\bfn^\pp,\bfm^\pp}
(x_1,x_2) = \la\scrO^{k,h}_{\bfel,\bfn,\bfm}(x_1)
\bar\scrO^{k^\pp,h^\pp}_{\bfel^\pp,\bfn^\pp,\bfm^\pp}(x_2)\ra \, ,
\label{gen2ptf}
\ee
where conformal invariance and conservation of $J$ require
$k+h=k^\pp+h^\pp$.

The combinatorics involved in the calculation of (\ref{gen2ptf}) is
rather formidable and we shall not present a detailed computation of
the complete two-point functions. However, our analysis will be
sufficient to determine the dependence of the two-point functions in
this class on the parameters $\gy$, $N$ and $J$, which will be
compared with the result of the dual string amplitude in section
\ref{string2pt}.

As previously observed, the only non-zero free fermion propagators are
$\la\psi^{-\,a}_\a\bar\psi^{-\,\bdot\dot b}\ra$ and
$\la\psi^{+\,\adot}_{\dot a} \bar\psi^+_{\b b}\ra$. This implies that
the two-point functions in the class (\ref{gen2ptf}) are only non-zero
at tree level if $k=k^\pp$ and $h=h^\pp$. We will now show that
instanton contributions to these correlation functions are non-zero,
in the leading semi-classical approximation, if the weaker condition
$k+h=k^\pp+h^\pp$ imposed by the symmetries is satisfied.

The dependence on the parameters, $\gy$, $N$ and $J$, in the two-point
function (\ref{gen2ptf}) can be determined analysing the structure of
the fermion zero modes in the classical profiles of the operators in
the instanton background. The combinations of scalar impurities
entering into $\scrO^{k,h}_{\bfel,\bfn,\bfm}$ and
$\bar\scrO^{k^\pp,h^\pp}_{\bfel^\pp,\bfn^\pp,\bfm^\pp}$ are
\ba
&& +\v^{12}\v^{13}\v^{24}\v^{34} \, , \quad
-\v^{12}\v^{34}\v^{24}\v^{13} \, , \quad
+\v^{12}\v^{24}\v^{34}\v^{13} \, , \nn \\
&& +\v^{12}\v^{34}\v^{13}\v^{24} \, , \quad
-\v^{12}\v^{13}\v^{34}\v^{24} \, , \quad
-\v^{12}\v^{24}\v^{13}\v^{34} \, ,
\label{scalimps}
\ea
and cyclic permutations of these. All the terms in (\ref{scalimps})
contain the same combination of fermion zero modes,
\be
\left(\scrmf^1\right)^2\left(\scrmf^2\right)^2
\left(\scrmf^3\right)^2\left(\scrmf^4\right)^2 \, ,
\label{scalfermodes}
\ee
where $\scrmf^A$ indicates a generic fermion zero mode in the
one-instanton sector, \ie either a superconformal mode, $\eta^A$ or
$\bar\xi^A$, or a mode of type $\nu^A$ or $\bar\nu^A$.

The zero modes contained in the pairs of fermionic impurities in
$\scrO^{k,h}_{\bfel,\bfn,\bfm}$ are
\ba
&& \veps_{ab}\,\psi^{-\,\a a}\psi_a^{-\,b}\sim\scrmf^1\scrmf^4 \nn \\
&& \veps^{\dot a\dot b} \, \psi^+_{\adot\dot a}\psi^{+\,\adot}_{\dot b}
\sim\left(\scrmf^1\right)^2\scrmf^2\scrmf^3\left(\scrmf^4\right)^2
\, . \label{ferfermodes}
\ea
Similarly the fermionic impurities in
$\bar\scrO^{k^\pp,h^\pp}_{\bfel^\pp,\bfn^\pp,\bfm^\pp}$ contain
\ba
&& \veps_{\dot a\dot b}\, \bar\psi^{-\,\dot a}_\adot
\bar\psi^{-\,\adot\dot b}\sim\scrmf^1\left(\scrmf^2\right)^2
\left(\scrmf^3\right)^2\scrmf^4 \nn \\
&& \veps^{ab}\,\bar\psi^{+\,\a}_a\bar\psi^+_{\a b} \sim
\scrmf^2\scrmf^3 \, .
\label{bferfermodes}
\ea
Taking into account the $J$ $Z$ fields in $\scrO$ and the $J$ $\barZ$
fields in $\bar\scrO$ the two operators contain the following
combinations of fermion modes
\ba
\what\scrO^{k,h}_{\bfel,\bfn,\bfm} &\to&
\left(\scrmf^1\right)^{J+2+k+2h}
\left(\scrmf^2\right)^{2+h} \left(\scrmf^3\right)^{2+h}
\left(\scrmf^4\right)^{J+2+k+2h} \nn \\
\what{\!\!\bar\scrO}^{k^\pp,h^\pp}_{\bfel^\pp,\bfn^\pp,\bfm^\pp}
&\to& \left(\scrmf^1\right)^{2+k^\pp}
\left(\scrmf^2\right)^{J+2+2k^\pp+h^\pp}
\left(\scrmf^3\right)^{J+2+2k^\pp+h^\pp}\left(\scrmf^4\right)^{2+k^\pp}
\, . \label{optotfermodes}
\ea
The computation of the two-point function
$G^{k,h;k^\pp,h^\pp}_{\bfel,\bfn,\bfm;\bfel^\pp,\bfn^\pp,\bfm^\pp}
(x_1,x_2)$ in the semi-classical approximation involves the
integration over the sixteen fermion superconformal modes associated
with the broken Poincar\'e and special supersymmetries. To saturate
these integrations the two operators must both contain
$\prod_{A=1}^4\left(\zeta^A\right)^2$, where $\zeta^A_\a =
\fr{\sqrt{\rho}}[\rho\eta^A_\a-(x-x_0)_\mu\s^\mu_{\a\adot}\bar\xi^{\adot
A}]$. This requirement combined with (\ref{optotfermodes}) implies
that in the product of the profiles of the two operators one must
select terms containing
\ba
&&\hsp{-1}\what\scrO^{k,h}_{\bfel,\bfn,\bfm}(x_1)\;\;
\what{\!\!\bar\scrO}^{k^\pp,h^\pp}_{\bfel^\pp,\bfn^\pp,\bfm^\pp}(x_2)
\to \left[\left(\zeta^1\right)^2\left(\zeta^2\right)^2
\left(\zeta^3\right)^2\left(\zeta^4\right)^2\right]\!(x_1)
\left[\left(\zeta^1\right)^2\left(\zeta^2\right)^2
\left(\zeta^3\right)^2\left(\zeta^4\right)^2\right]\!(x_2) \nn \\
&&\hsp{-1}\times \left(\nu^1+\bar\nu^1\right)^{J+k+2h+k^\pp}
\left(\nu^2+\bar\nu^2\right)^{J+h+2k^\pp+h^\pp}
\left(\nu^3+\bar\nu^3\right)^{J+h+2k^\pp+h^\pp}
\left(\nu^4+\bar\nu^4\right)^{J+k+2h+k^\pp}  ,
\label{genprofiles}
\ea
where the $\nu$ and $\bar\nu$ modes will eventually be paired in
colour singlet bilinears.

As discussed in \cite{gks2} the integration over the five-sphere
imposes the condition that $\nu$ and $\bar\nu$ modes of each flavour
appear with the same multiplicity. From (\ref{genprofiles}) we thus
get the condition
\be
J+k+2h+k^\pp = J+h+2k^\pp+h^\pp \; \Rightarrow
\; k+h=k^\pp+h^\pp \, ,
\label{5spherecond}
\ee
which is automatically satisfied by all the two-point functions
allowed by the symmetries.

Equation (\ref{genprofiles}) is the starting point to study the
dependence of the two-point function (\ref{gen2ptf}) on the parameters
$\gy$, $N$ and $J$. In the profile of the operator
$\scrO^{k,h}_{\bfel,\bfn,\bfm}$ the superconformal modes of flavour 2
and 3 can only be taken from the impurities whereas the modes of
flavour 1 and 4 can come either from the impurities or from the
$Z$'s. As in the examples discussed in \cite{gks2} the dominant
contributions in the large $J$ limit come from terms in which all the
$\zeta^1$'s and $\zeta^4$'s are provided the $Z$'s because in this
case a factor of $J$ is associated with the choice of each $Z$
providing one such mode.

Satisfying the condition (\ref{5spherecond}) is not sufficient to
ensure that the two-point function (\ref{gen2ptf}) receives a non zero
instanton contribution in the BMN limit. In order to cancel the
inverse powers of $N$ coming from the normalisation of the operators
it is necessary to combine all the $\nu$ and $\bar\nu$ modes in
antisymmetric bilinears, $\nsix$. In the two and four impurity cases
studied in \cite{gks2} this requirement was always satisfied. In the
case of the operators under consideration the requirement is
non-trivial and has important consequences. The traces in the
definition of the operators can be explicitly evaluated using the form
of the instanton solution for the elementary fields. In particular,
the solution for the anti-chiral fermions $\bar\lambda_A^\adot$, whose
components $\psi^+$ and $\bar\psi^-$ enter
$\scrO^{k,h}_{\bfel,\bfn,\bfm}$ and
$\bar\scrO^{k^\pp,h^\pp}_{\bfel^\pp,\bfn^\pp,\bfm^\pp}$ respectively,
was given in \cite{dhkm}. Selecting in such traces the terms which
contain the correct combinations of superconformal modes shows that if
any $Z$'s are inserted between two contracted $\psi^+$'s it is not
possible to antisymmetrise all the $\bar\nu\nu$ bilinears, because
necessarily colour contractions between a $\nu$ and a $\bar\nu$ of the
same flavour occur. Such contributions are suppressed at large $N$
(see equation (\ref{largeNdep})) and vanish in the BMN limit. This
means that non-vanishing contributions in the BMN limit come only from
the terms with $v_{2i}=0$, $i=1,\ldots,h$ in the sums in
(\ref{genBMNop}), effectively reducing the number of sums involved in
the calculation of the operator profile. Analogously in the operator
$\bar\scrO^{k^\pp,h^\pp}_{\bfel^\pp,\bfn^\pp,\bfm^\pp}$ no $\barZ$'s
can be inserted between two contracted $\bar\psi^-$'s implying the
constraint $u^\pp_{2i}=0$, $i=1,\dots,k^\pp$. This observation is
crucial in determining the $J$ dependence of the two-point functions
we are considering, notably in proving that they obey BMN scaling.

In all the relevant contributions to the profile of the operators
(\ref{genBMNop}) the traces are independent of the way the $Z$'s are
grouped and only depend on the relative order of the impurities, \ie
they do not depend on the summation indices $p$, $q$, $r$, $u_1$,
\ldots, $v_{2h}$. All the traces in
$\what\scrO^{k,h}_{\bfel,\bfn,\bfm}$ that contribute in the BMN limit
can be reduced to the form
\ba
\what\scrO^{k,h}_{\bfel,\bfn,\bfm} &\to&
\frac{\rho^{8+2k-2h}\,(x-x_0)^{2h}}{[(x-x_0)^2+\rho^2]^{J+3k+3h+8}}
\left[\left(\zeta^1\right)^2\left(\zeta^2\right)^2
\left(\zeta^3\right)^2\left(\zeta^4\right)^2\right]
\left(\bar\nu^{[1}\nu^{4]}\right)^{J+k+h} \nn \\
&& \times \left[c_1\left(\bar\nu^{[1}\nu^{2]}\right)
\left(\bar\nu^{[3}\nu^{4]}\right) +
c_2\left(\bar\nu^{[1}\nu^{3]}\right)
\left(\bar\nu^{[2}\nu^{4]}\right) \right]^h \, .
\label{opprof}
\ea
Similarly all the relevant traces in the profile of the conjugate
operator reduce to
\ba
\what{\!\!\bar\scrO}^{k^\pp,h^\pp}_{\bfel^\pp,\bfn^\pp,\bfm^\pp}
&\to& \frac{\rho^{8-2k^\pp+2h^\pp}\,(x-x_0)^{2k^\pp}}
{[(x-x_0)^2+\rho^2]^{J+3k^\pp+3h^\pp+8}}
\left[\left(\zeta^1\right)^2\left(\zeta^2\right)^2
\left(\zeta^3\right)^2\left(\zeta^4\right)^2\right]
\left(\bar\nu^{[2}\nu^{3]}\right)^{J+k^\pp+h^\pp} \nn \\
&& \times \left[ c_1^\pp \left(\bar\nu^{[1}\nu^{3]}\right)
\left(\bar\nu^{[2}\nu^{4]}\right)+
c_2^\pp \left(\bar\nu^{[1}\nu^{2]}\right)
\left(\bar\nu^{[3}\nu^{4]}\right)\right]^{k^\pp} \, .
\label{bopprof}
\ea
In (\ref{opprof}) and (\ref{bopprof}) $c_1$, $c_2$, $c_1^\pp$ and
$c_2^\pp$ denote numerical coefficients. As in the cases studied in
\cite{gks2} the dependence on the summation indices is thus only in
the phases and in combinatorial factors associated with the
multiplicity of each contribution. The traces (\ref{opprof}) and
(\ref{bopprof}) can be factored out of the sums. This simplifies the
calculation and especially the analysis of the $J$ dependence.

The definition of $\scrO^{k,h}_{\bfel,\bfn,\bfm}$ involves a sum over
$3+2k+2h$ indices. However, as observed above, the number of sums is
reduced by the requirement that all the $\bar\nu\nu$ bilinears be
antisymmetrised, which implies that no $Z$'s can be inserted between
two contracted $\psi^+$ fermions. Hence effectively the classical
profile of the operator $\scrO^{k,h}_{\bfel,\bfn,\bfm}$ contains only
$3+2k+h$ sums. Similarly the profile of
$\bar\scrO^{k^\pp,h^\pp}_{\bfel^\pp,\bfn^\pp,\bfm^\pp}$ contains only
$3+k^\pp+2h^\pp$ sums.  Taking into account the multiplicity factors
associated with the choice of the four $Z$'s and the four
$\barZ$'s which provide respectively the superconformal modes of
flavour 1 and 4 in $\scrO^{k,h}_{\bfel,\bfn,\bfm}$ and  those of
flavour 2 and 3 in
$\bar\scrO^{k^\pp,h^\pp}_{\bfel^\pp,\bfn^\pp,\bfm^\pp}$, the sums in
$\scrO^{k,h}_{\bfel,\bfn,\bfm}$ contribute to the two-point function a
factor of $J^{7+2k+h}$ and those in
$\bar\scrO^{k^\pp,h^\pp}_{\bfel^\pp,\bfn^\pp,\bfm^\pp}$ a factor of
$J^{7+k^\pp+2h^\pp}$. For instance choosing all the four $Z$'s
from the second group of $p_2$ $Z$'s in the trace in (\ref{genBMNop})
leads to the sum
\be
\hspace*{-0.3cm} \begin{array}[t]{c}
{\displaystyle \sum_{p_2,p_3,p_4,u_1,\dots,v_{2h-1}=0}^J} \\
{\scriptstyle p_2+p_3+p_4+u_1+\cdots+v_{2h-1}\le J} \\
{\scriptstyle p_1=J-(p_2+p_3+p_4+u_1+\cdots+v_{2h-1})}
\end{array} \hspace*{-0.5cm}
e(\bfp,\bfu,\bfv;\bfel,\bfn,\bfm;J)
\,\fr{4!}\,p_2(p_2-1)(p_2-2)(p_2-3)
\sim J^{7+2k+h} \, ,
\label{sumJ}
\ee
where only the $v_i$'s with odd index are summed over. The
combinatorics associated with these sums becomes increasingly involved
as the number of impurities grows. In the case of the four impurity
operators of \cite{gks2} there were 35 independent traces to
compute. In the general case of the operator (\ref{genBMNop}) for a
fixed relative order of the impurities the number of independent
traces associated with the choice of the four $Z$'s which soak up
superconformal modes is
\be
\frac{(2k+h+7)!}{4!(2k+h+3)!} \,
\label{traceno}
\ee
and moreover one has to sum the contributions corresponding to the
different relative orderings of the impurities, since the operators
considered here, unlike those of \cite{gks2}, involve impurities of
different types, bosonic and fermionic ones.  The sums such as
(\ref{sumJ}) also encode the dependence of the operator profiles on
the integers in $\bfel$, $\bfn$ and $\bfm$, corresponding to
the mode numbers of the dual string states. Each of the sums
contributing to any operator in this class gives rise to a very
complicated dependence on the mode numbers. We shall see, however,
that the string theory analysis predicts a very simple dependence,
requiring dramatic simplifications on the gauge theory side.

As in the cases considered in \cite{gks2}, the other elements which
determine the dependence on the parameters $\gy$, $N$ and $J$ are,
apart from the normalisation of the operators, the number of $\nsix$
bilinears, the bosonic integrals over $x_0$ and $\rho$ and the
five-sphere integrals.

Equations (\ref{opprof}) and (\ref{bopprof}) show that the profiles of
the two operators contain a total of $2J+k+3h+3k^\pp+h^\pp$ $\nsix$
bilinears, each producing a factor of $\gy\sqrt{N}$, so that the total
contribution to the two-point function of the $\nsix$ bilinears is
$(\gy\sqrt{N})^{2J+k+3h+3k^\pp+h^\pp}$.

The integrations over $x_0$ and $\rho$ are logarithmically divergent
and need to be regularised, \eg by dimensional regularisation of the
$x_0$ integral. They can then be evaluated using standard techniques
and are found to behave as $1/J^2$ in the large $J$ limit.

Finally, additional powers of $J$ arise from the five-sphere
integration after re-expressing the $\nsix$ bilinears in terms of
$\Omega^{AB}$'s \cite{gks2}. The combinations of
$\bar\nu^{[A}\nu^{B]}$ bilinears to consider are those in
(\ref{opprof}) and (\ref{bopprof}). The resulting five sphere
integrals are all of the form
\be
I_{S^5}(k,h;k^\pp,h^\pp) = \int \dr^5\Omega \,
\left(\Omega^{14}\right)^{J+k+h}
\left(\Omega^{23}\right)^{J+k^\pp+h^\pp}
\left[\left(\Omega^{12}\right) \left(\Omega^{34}\right)
+\left(\Omega^{13}\right) \left(\Omega^{24}\right)\right]^{h+k^\pp}
\, . \label{RR5sph}
\ee
Using the constraint $k+h=k^\pp+h^\pp$ these integrals can be put in
the form (\ref{gen5sph}) with $a=J+k+h=J+k^\pp+h^\pp$ and
$b+c=h+k^\pp$. Therefore (\ref{5sphabc-res}) immediately gives
\be
I_{S^5}(k,h;k^\pp,h^\pp)\big|_{k+h=k^\pp+h^\pp}
\sim \fr{J^{2+h+k^\pp}} \, .
\label{5sph-Jdep}
\ee

Combining the various contributions described above with the
normalisation factors and the moduli space measure, we can summarise
the dependence on $\gy$, $N$ and $J$ in
$G^{k,h;k^\pp,h^\pp}_{\bfel,\bfn,\bfm;\bfel^\pp,\bfn^\pp,\bfm^\pp}(x_1,x_2)$
as follows
\ba
&& \underbrace{\fr{\sqrt{J^{3+2k+2h}(\gy^2N)^{J+4+2k+2h}}}
\fr{\sqrt{J^{3+2k^\pp+2h^\pp}(\gy^2N)^{J+4+2k^\pp+2h^\pp}}}}_{\rm
normalised ~ op.~ profiles}\;\underbrace{\er^{2\pi i\tau}
\gy^8\sqrt{N}}_{\rm measure} \nn \\
&& \underbrace{\left(\gy\sqrt{N}\right)^{2J+k+3h+3k^\pp+h^\pp}}_{\nu,
\:\bar\nu ~ {\rm bilinears}}\;
\underbrace{\fr{J^2}}_{x_0,\:\rho ~ {\rm integrals}}
\;\;\underbrace{\fr{J^{2+h+k^\pp}}}_{S^5 ~ {\rm integral}}
\;\underbrace{J^{7+2k+h}J^{7+k^\pp+2h^\pp}}_{\rm sums} \nn \\
&& = J^{7+k-h-k^\pp+h^\pp}\,\gy^{-k+h+k^\pp-h^\pp}\,
N^{-\frac{7}{2}+\half(-k+h+k^\pp-h^\pp)}\,\er^{2\pi i\tau}
\rule{0pt}{16pt} \nn \\
&& = \left(\frac{J^2}{N}\right)^{7/2}
\left(\frac{J^2}{\gy^2N}\right)^{\half(k-h-k^\pp+h^\pp)}
\er^{2\pi i\tau}
\label{gNJdep} \, ,
\ea
so that the behaviour of the generic two-point functions in this class
is
\be
G^{k,h;k^\pp,h^\pp}_{\bfel,\bfn,\bfm;\bfel^\pp,\bfn^\pp,\bfm^\pp}
(x_1,x_2) \sim \frac{\left(g_2\right)^{7/2}}
{\left(\lambda^\pp\right)^{\half(k-h-k^\pp+h^\pp)}}
\,\er^{-\frac{8\pi^2}{g_2\lambda^\pp}+i\theta} \, .
\label{gNJdepfin}
\ee
The first thing to notice is that (\ref{gNJdepfin}) shows that the
two-point functions respect BMN scaling. The leading instanton
contribution can be re-expressed in terms of the parameters
$\lambda^\pp$ and $g_2$. The arguments given in \cite{gks2} to
illustrate how the subleading corrections can give rise to a double
series in $\lambda^\pp$ and $g_2$ can be repeated in the present
case. Therefore one can argue that the BMN scaling property of
(\ref{gen2ptf}) extends beyond the semi-classical approximation.

It is interesting to consider special cases of (\ref{gNJdepfin}). If
$k$, $h$, $k^\pp$ and $h^\pp$ are chosen in such a way that the
two-point function is non-zero at tree level, \ie $k=k^\pp$ and
$h=h^\pp$, the leading instanton contribution has no powers of
$\lambda^\pp$. This is the same behaviour found for the four impurity
singlet operators.

In general instanton corrections to two-point functions which vanish
at tree level start with a non-zero power of $\lambda^\pp$.
Interestingly, among these there is a class of two-point functions for
which the leading non-zero contribution contains negative powers of
$\lambda^\pp$. The simplest examples of this type involve the
operators $\scrO^{k,0}$, with only $\psi^-$ insertions, and
$\bar\scrO^{0,h^\pp}$, with only $\bar\psi^+$ insertions. Notice,
however, that although two-point functions of this type can have
arbitrarily large powers of $\lambda^\pp$ in the denominator, they are
not singular in the $\lambda^\pp\to 0$ limit because of the
exponential factor $\exp(-8\pi^2/\lambda^\pp g_2)$.

\subsection{A class of mixed \RR\!\!--\NSNS two-point functions}
\label{RRNSNS2pt}

We now study another class of correlation functions which vanish at
tree level but receive instanton contributions, namely two-point
functions corresponding to string amplitudes mixing \RR and \NSNS
states. The general two-point function we consider is
\be
G^{k,h;l}_{\bfel,\bfn,\bfm;\bfel^\pp,\bfn^\pp}(x_1,x_2) =
\la \scrO^{k,h}_{\bfel,\bfn,\bfm}(x_1)\,
\bar\scrO^l_{\bfel^\pp,\bfn^\pp}(x_2)\ra \, ,
\label{genmix2ptf}
\ee
where $\scrO^{k,h}_{\bfel,\bfn,\bfm}$ is an operator with fermionic
impurities of the form (\ref{genBMNop}) and
$\bar\scrO^l_{\bfel^\pp,\bfn^\pp}$ is the conjugate of the operator
defined in (\ref{genBMNopNS}). Conformal invariance and the U(1)
symmetry impose in this case the constraint $l=k+h$.

Much of the analysis in the previous subsection can be applied to
(\ref{genmix2ptf}). The contribution of the profile of
$\scrO^{k,h}_{\bfel,\bfn,\bfm}$ is the same and we only need to
discuss the \NSNS operator $\bar\scrO^l_{\bfel^\pp,\bfn^\pp}$. The
classical expression for $\scrO^{k,h}_{\bfel,\bfn,\bfm}$ contains the
combination of fermion modes in the first line of
(\ref{optotfermodes}),
\be
\left(\scrmf^1\right)^{J+2+k+2h}
\left(\scrmf^2\right)^{2+h} \left(\scrmf^3\right)^{2+h}
\left(\scrmf^4\right)^{J+2+k+2h} \, .
\label{rropfermodes}
\ee
In order to get a non-zero contribution to the two-point function
(\ref{genmix2ptf}) we need to select terms in
$\bar\scrO^l_{\bfel^\pp,\bfn^\pp}$ in which the impurities contain
fermion modes of each flavour with the same multiplicity. This means
that in $\;\what{\!\!\bar\scrO}^{\,l}_{\!\!\bfel^\pp,\bfn^\pp}$ we keep
terms containing
\be
\left(\scrmf^1\right)^{l+2}
\left(\scrmf^2\right)^{J+2+l} \left(\scrmf^3\right)^{J+2+l}
\left(\scrmf^4\right)^{l+2} \, .
\label{nsopfermodes}
\ee
The double scaling limit, $N\to\infty$, $J\to\infty$, with $J^2/N$
finite, requires that once the fermion superconformal modes are soaked
up, all the modes of type $\nu$ and $\bar\nu$ be combined in $\nsix$
bilinears. All the relevant terms in the profiles of the operators
$\scrO^{k,h}_{\bfel,\bfn,\bfm}$ and $\bar\scrO^l_{\bfel^\pp,\bfn^\pp}$
can then be reduced to the form
\ba
\what\scrO^{k,h}_{\bfel,\bfn,\bfm} &\to&
\frac{\rho^{8+2k-2h}\,(x-x_0)^{2h}}{[(x-x_0)^2+\rho^2]^{J+3k+3h+8}}
\left[\left(\zeta^1\right)^2\left(\zeta^2\right)^2
\left(\zeta^3\right)^2\left(\zeta^4\right)^2\right]
\left(\bar\nu^{[1}\nu^{4]}\right)^{J+k+h} \nn \\
&& \times \left[c_1\left(\bar\nu^{[1}\nu^{2]}\right)
\left(\bar\nu^{[3}\nu^{4]}\right) +
c_2\left(\bar\nu^{[1}\nu^{3]}\right)
\left(\bar\nu^{[2}\nu^{4]}\right) \right]^h \, ,
\label{rropprof} \\
\what{\!\!\bar\scrO}^{\,l}_{\!\bfel^\pp,\bfn^\pp} &\to&
\frac{\rho^{8}}{[(x-x_0)^2+\rho^2]^{J+3l+8}}
\left[\left(\zeta^1\right)^2\left(\zeta^2\right)^2
\left(\zeta^3\right)^2\left(\zeta^4\right)^2\right]
\left(\bar\nu^{[2}\nu^{3]}\right)^{J+l} \nn \\
&& \times \left[c^\pp_1\left(\bar\nu^{[1}\nu^{2]}\right)
\left(\bar\nu^{[3}\nu^{4]}\right) +
c^\pp_2\left(\bar\nu^{[1}\nu^{3]}\right)
\left(\bar\nu^{[2}\nu^{4]}\right) \right]^l \, ,
\label{nsopprof}
\ea
where $c_1$, $c_2$, $c^\pp_1$ and $c^\pp_2$ are numerical
coefficients.

We can now repeat the analysis of the previous subsection to determine
the dependence of (\ref{genmix2ptf}) on $\gy$, $N$ an $J$. As in the
case of the \RR two-point function the terms of the form
(\ref{rropprof}) in $\what\scrO^{k,h}_{\bfel,\bfn,\bfm}$, which
contribute in the BMN limit, involve only $3+2k+h$ sums, so the
resulting contribution is a factor of $J^{7+2k+h}$. In the operator
$\bar\scrO^l_{\bfel^\pp,\bfn^\pp}$ there is no restriction on the
sums, which therefore contribute a factor of $J^{7+2l}$.

The total number of $\nsix$ bilinears in the two-point function is
$2J+k+3h+3l$, so that the resulting contribution is
$(\gy\sqrt{N})^{2J+k+3h+3l}$.

The $x_0$ and $\rho$ integrals are logarithmically divergent and after
regularisation can be shown to behave as $1/J^2$ in the BMN limit.

The five-sphere integrals are again of the form (\ref{gen5sph}). From
(\ref{rropprof}) and (\ref{nsopprof}) we get
\be
\int \dr^5\Omega \, \left(\Omega^{14}\right)^{J+k+h}
\left(\Omega^{23}\right)^{J+l}
\left[\left(\Omega^{12}\right) \left(\Omega^{34}\right)
+\left(\Omega^{13}\right) \left(\Omega^{24}\right)\right]^{h+l}
\, , \label{mix5sph}
\ee
which according to the general formula (\ref{5sphabc-res}) behaves as
$1/J^{2+h+l}$.

Combining all these contributions we can determine the behaviour of
the two-point function (\ref{genmix2ptf}) in the BMN limit,
\ba
&& \underbrace{\fr{\sqrt{J^{3+2k+2h}(\gy^2N)^{J+4+2k+2h}}}
\fr{\sqrt{J^{3+2l}(\gy^2N)^{J+4+3l}}}}_{\rm
normalised ~ op.~ profiles}\;\underbrace{\er^{2\pi i\tau}
\gy^8\sqrt{N}}_{\rm measure} \nn \\
&& \underbrace{\left(\gy\sqrt{N}\right)^{2J+k+3h+3l}}_{\nu,
\:\bar\nu ~ {\rm bilinears}}\;
\underbrace{\fr{J^2}}_{x_0,\:\rho ~ {\rm integrals}}
\;\;\underbrace{\fr{J^{2+h+l}}}_{S^5 ~ {\rm integral}}
\;\underbrace{J^{7+2k+h}J^{7+2l}}_{\rm sums} \nn \\
&& = J^{7+k-h}\,\gy^{-k+h}\,
N^{-\frac{7}{2}-\half(k-h)}\,\er^{2\pi i\tau}
\rule{0pt}{16pt} \nn \\
&& = \left(\frac{J^2}{N}\right)^{7/2}
\left(\frac{J^2}{\gy^2N}\right)^{\half(k-h)}
\er^{2\pi i\tau}
\label{mixgNJdep} \, .
\ea
The result for the generic two-point function in this class is thus
\be
G^{k,h;l}_{\bfel,\bfn,\bfm;\bfel^\pp,\bfn^\pp}(x_1,x_2)
\sim \frac{(g_2)^{7/2}\,\er^{-\frac{8\pi^2}{g_2\lambda^\pp}+i\theta}}
{(\lambda^\pp)^{\half(k-h)}}
\, . \label{mixgNJdepfin}
\ee
This shows that mixed \RR\!\!--\NSNS correlation functions of this
type receive a non-zero contribution in the one-instanton sector in
the BMN limit, if the condition $l=k+h$ required by the symmetries is
satisfied. The result (\ref{mixgNJdepfin}) respects BMN scaling. As in
the case considered in the previous subsection, depending on the
number of $\psi^-$ and $\psi^+$ impurities, the leading contribution
can start with a positive of negative (half integer) power of
$\lambda^\pp$.

Before considering the dual string calculation, we conclude this
section with a small digression. The previous analysis allows to
easily determine the behaviour of the leading instanton contribution
to two-point functions of singlet operators with an arbitrary number
of scalar impurities. In \cite{gks2} it was shown that for two
impurity operators the leading instanton contribution to the anomalous
dimension is
\be
\g_{\rm 2-impur}\sim \left(g_2\right)^{7/2}\left(\lambda^\pp\right)^2
\er^{2\pi i\tau} \, ,
\label{2impres}
\ee
whereas four impurity operators receive a leading contribution of
order
\be
\g_{\rm 4-impur}\sim \left(g_2\right)^{7/2}\er^{2\pi i\tau} \, .
\label{4impres}
\ee
Repeating step by step the calculations in this section shows that the
two-point function
\be
G^l(x_1,x_2) = \la \scrO^l(x_1) \, \bar\scrO^l(x_2) \ra \, ,
\label{gennsns2ptf}
\ee
where $\scrO^l$ is of the form (\ref{genBMNopNS}), behaves as
$(g_2)^{7/2}\,\er^{2\pi i\tau}$. Therefore in general operators with
only scalar impurities, at least in the class of singlets we are
considering, are expected to have an instanton induced anomalous
dimension
\be
\g^{\rm (1-inst)} \sim \left(g_2\right)^{7/2}\er^{2\pi i\tau} \, ,
\label{genscaladim}
\ee
irrespective of the number of impurities.

In the next section we will show that the dual string amplitudes
precisely reproduce all the features of the gauge theory two-point
functions discussed here and in section \ref{RR2pt}. We will also see
that string theory predicts a very simple result for the mode number
dependence, which is extremely complicated to extract from a gauge
theory calculation.

\section{Plane-wave string two-point amplitudes}
\label{string2pt}

\subsection{$D$-instanton induced two-point amplitudes}
\label{Dgeneral}

The two-point functions discussed in the previous section are dual to
$D$-instanton induced plane-wave string scattering amplitudes between
external states of the form (\ref{genstringstate}) and
(\ref{genstrstatens}). $D$-instanton contributions to such amplitudes
are computed using the boundary state constructed in \cite{gg} and the
formalism of \cite{gks1}.

The leading $D$-instanton contribution to two-point amplitudes comes
from diagrams in which the external states are coupled to two separate
disks,
\be
\scrA_{\bfr,\bfs} = \gs^{7/2}\,\er^{2\pi i\tau}\,
{}_{_1}\!\la\chi_{\bfr}|\otimes{}_{_2}\!\la\chi_{\bfs}
\|V_2\ra\!\ra \, ,
\label{genstrampl}
\ee
where the prefactor, $\gs^{7/2}\,\er^{2\pi i\tau}$, is (up to
a numerical constant) the measure on the single $D$-instanton moduli
space and $\tau=C^{\scriptscriptstyle(0)}+i\er^{-\phi}$, where
$C^{\scriptscriptstyle(0)}$ is the \RR scalar and $\phi$ the dilaton.
In (\ref{genstrampl}) $|\chi_{\bfr}\ra_{_1}$ and
$|\chi_{\bfs}\ra_{_2}$ denote the incoming and outgoing states
respectively and $\bfr$ and $\bfs$ collectively indicate the
corresponding quantum numbers, including the mode
numbers. $\|V_2\ra\!\ra$ is the dressed two-boundary state
\cite{gks1}, which contains the dependence on the bosonic and
fermionic moduli and couples to any pair of physical states,
\be
\|V_2\ra\!\ra = \d^4(\bar\theta_{2L}+\bar\theta_{1L})\,
\d^4(\bar\theta_{2R}+\bar\theta_{1R})\int\dr^8\eta\left[\eta
(Q^-_1+Q^-_2)\right]^8\,\|\hat V_2^{(0)}\ra\!\ra \, ,
\label{dressbs1}
\ee
where
\ba
\|\hat V_2^{(0)}\ra\!\ra &=& (2\pi)^8
\exp\left[ \sum_{k=1}^{\infty}
\frac{1}{\omega_k} \alpha^{(1)I}_{-k} \tilde\alpha^{(1) I}_{-k}
- i  S^{(1)}_{-k}M_k\tilde{S}^{(1)}_{-k}\right. \nn \\
&& \hsp{2.1} \left. + \frac{1}{\omega_k} \alpha^{(2)I}_{-k}
\tilde\alpha^{(2)I}_{-k}- i  S^{(2)}_{-k}M_k\tilde{S}^{(2)}_{-k}
\right] \er^{-{\bfa^\dagger_ 1}\cdot {\bfa^\dagger_2}} \,
|0 \rangle_{_1}\! \otimes |0 \rangle_{_2} \, .
\label{zinnt}
\ea
In (\ref{dressbs1}) $Q^-_1$ and $Q^-_2$ denote the broken dynamical
supersymmetries on the two disks and in (\ref{zinnt})
$\er^{-{\bfa^\dagger_1} \cdot {\bfa^\dagger_2}} |0
\rangle_{_1}\! \otimes  |0 \rangle_{_2}$ is the zero-mode part of the
two-boundary state after integration over the transverse position
moduli. The $\d$-functions in (\ref{dressbs1}) arise after integration
over the fermion moduli associated with the broken kinematical
supersymmetries.

The relations (\ref{param-ids}) between the string and gauge theory
parameters imply that   in order to make contact with the
semi-classical calculations of the previous section in the double
scaling limit, $J\to\infty$, $N\to\infty$, with $J^2/N$ fixed, we need
to study the relevant string amplitudes (\ref{genstrampl}) in the
small $g_s$ and large $m$ limit.

In computing amplitudes such as (\ref{genstrampl}) one expands the
dressed two-boundary state, $\|V_2\ra\!\ra$, retaining only the terms
which, commuted through the eight dynamical supercharges, give a
non-zero result acting to the left as annihilation operators on the
external states. The large $m$ limit, relevant for the comparison with
the gauge theory, selects very specific contributions in this
expansion.

\subsection{Amplitudes in the \RR sector}
\label{RRstring}

To make contact with the calculation of the two-point functions in
section \ref{RR2pt} we are interested in amplitudes such as
(\ref{genstrampl}), where the external states are of the form
(\ref{genstringstate}). So we consider
\ba
{}_{_1}\!\la\chi^{k,h}_{\bfel,\bfn,\bfm}|\otimes
{}_{_2}\!\la\chi^{k^\pp,h^\pp}_{\bfel^\pp,\bfn^\pp,\bfm^\pp}|
&=& \veps_{i_1i_2i_3i_4}\,\veps_{j_1j_2j_3j_4} \, \fr{\omega_{\ell_1}
\omega_{\ell_2}\omega_{\ell_1^\pp}\omega_{\ell_2^\pp}}
\label{extstate} \\
&\times& {}_h\!\la0|\a^{(1)i_1}_{\ell_1}\a^{(1)i_2}_{\ell_2}
\tilde\a^{(1)i_3}_{\ell_1}\tilde\a^{(1)i_4}_{\ell_2} \prod_{r=1}^k
\left[S_{n_r}^{(1)-}\tilde S_{n_r}^{(1)-} \right] \prod_{s=1}^h
\left[S_{m_s}^{(1)+}\tilde S_{m_s}^{(1)+} \right] \nn \\
&\otimes& {}_h\!\la0|\a^{(2)j_1}_{\ell_1^\pp}\a^{(2)j_2}_{\ell_2^\pp}
\tilde\a^{(2)j_3}_{\ell_1^\pp}\tilde\a^{(2)j_4}_{\ell_2^\pp}
\prod_{r=1}^{k^\pp} \left[S_{n^\pp_r}^{(2)-}\tilde S_{n^\pp_r}^{(2)-}
\right] \prod_{s=1}^{h^\pp}\left[S_{m^\pp_s}^{(2)+}
\tilde S_{m^\pp_s}^{(2)+} \right] \, , \nn
\ea
where the square brackets indicate contraction of the spinor indices
in the two SO(4) factors via $\veps$ tensors and we have used the
same vector notation for the indices as in section
\ref{bmnops}. Equation (\ref{extstate}) includes the normalisation
factors for the states which had been omitted in
(\ref{genstringstate}).

In order to compare the results with the gauge theory semi-classical
approximation we consider the large $m$ limit in the amplitude
\be
\scrA^{k,h;k^\pp,h^\pp}_{\bfel,\bfn,\bfm;\bfel^\pp,\bfn^\pp,\bfm^\pp}
= \gs^{7/2}\er^{2\pi i\tau}\, {}_{_1}\!\la\chi^{k,h}_{\bfel,\bfn,\bfm}|
\otimes{}_{_2}\!\la\chi^{k^\pp,h^\pp}_{\bfel^\pp,\bfn^\pp,\bfm^\pp}
\|V_2\ra\!\ra \, .
\label{rrstrampl}
\ee
The analysis of the leading contributions in this limit follows
closely the one presented in \cite{gks1} for two and four impurity
operators. We first consider the bosonic oscillators in
(\ref{extstate}) which act to the right as annihilation operators on
the boundary state. These are compensated, as in the four impurity
case of \cite{gks1}, by lowering from the exponent in (\ref{zinnt})
two $SM\tilde S$ bilinears for each disk and commuting them through
the broken dynamical supersymmetries (four of which are distributed on
each disk in (\ref{dressbs1})) to obtain bosonic creation
operators. Recalling that in the large $m$ limit
\be
S_{-r} M_r\tS_{-r} \approx \frac{2m}{r} \, S_{-r}^- \tS_{-r}^-
+\frac{r}{2m} \, S^+_{-r} \tS^+_{-r}
\label{smsexp}
\ee
and using the commutation relations in the plane-wave background
\cite{met}, the annihilation of the bosonic oscillators contributes to
the amplitude a factor
\be
\frac{m^{12}}{\ell_1\ell_2\ell_1^\pp\ell_2^\pp} \, .
\label{boscontr}
\ee
The analysis of the contribution of the fermionic oscillators is then
straightforward. The only subtlety is related to the sign of $m$. In
our conventions the momenta of incoming states are positive and those
of outgoing states are negative, therefore $m=\mu\al p_->0$ on disk 1
and $m<0$ on disk 2. We need to expand the boundary state retaining
only the terms with $k+h$ fermionic bilinears on the first disk and
$k^\pp+h^\pp$ on the second disk in order to annihilate the factors in
the last two lines of (\ref{extstate}). The expansion (\ref{smsexp}),
valid for $m>0$, shows that on the first disk a $[S^-_{-r}\tS^-_{-r}]$
bilinear contributes a factor of $2m/r$, whereas a
$[S^+_{-r}\tS^+_{-r}]$ bilinear contributes a factor of $r/2m$. The
situation is reversed on the second disk. The parameter $m$ is
negative and as a result the coefficients of the two terms in the
expansion (\ref{smsexp}) are interchanged. We get a factor of
$r^\pp/2m$ for each $[S^-_{-r^\pp}\tS^-_{-r^\pp}]$ bilinear and a
factor of $2m/r^\pp$ for each $[S^+_{-r^\pp}\tS^+_{-r^\pp}]$ bilinear
in the outgoing state $|\chi^{k^\pp,h^\pp}_{\bfel^\pp,\bfn^\pp,
\bfm^\pp}\ra_{_2}$.

Combining all the contributions and taking into account the
normalisation of the external states we find that the leading
$D$-instanton contribution to the amplitude (\ref{rrstrampl}) is
\be
\scrA^{k,h;k^\pp,h^\pp}_{\bfel,\bfn,\bfm;\bfel^\pp,\bfn^\pp,\bfm^\pp}
\sim \gs^{7/2}\,\er^{2\pi i\tau} \,
m^{8+(k-h)+(h^\pp-k^\pp)} \,\fr{\ell_1\ell_2\ell_1^\pp\ell_2^\pp}\,
\frac{\prod_{i=1}^{k^\pp} n^\pp_i \, \prod_{j=1}^h m_j}
{\prod_{i=1}^k n_i \, \prod_{j=1}^{h^\pp} m^\pp_j} \, .
\label{strresult}
\ee
As in the cases studied in \cite{gks1} the $D$-instanton induced
amplitude is non-zero only if the mode numbers in both external states
are pairwise equal. Integration over the modulus corresponding to the
position of the $D$-instanton in the $x^+$ direction imposes energy
conservation in the amplitude. This further constrains the mode
numbers imposing that they be equal in pairs between the incoming and
outgoing state. However, in the large $m$ limit this condition reduces
to the requirement that the external states contain the same number of
oscillators.

The amplitude (\ref{strresult}) induces a correction to the string
mass matrix which, expressed in terms of Yang--Mills parameters and
rescaled by a factor of $\mu$, becomes
\be
\fr{\mu} \,\d M \sim g_2^{7/2}\,
\er^{-\frac{8\pi^2}{g_2\lambda^\pp}+i\theta} \,
\fr{(\lambda^\pp)^{\half(k-h+h^\pp-k^\pp)}} \,
\fr{\ell_1\ell_2\ell_1^\pp\ell_2^\pp}\,
\frac{\prod_{i=1}^{k^\pp} n^\pp_i \, \prod_{j=1}^h m_j}
{\prod_{i=1}^k n_i \, \prod_{j=1}^{h^\pp} m^\pp_j} \, .
\label{masscorr}
\ee
The dependence on the parameters, $g_2$ and $\lambda^\pp$, in this
result is in agreement with what we found in the dual Yang--Mills
correlation functions in section \ref{2ptfuncts}, equation
(\ref{gNJdepfin}). Moreover the string result shows a very simple
dependence on the mode numbers of the external states. On the other
hand, as already observed, the computation of the mode number
dependence in the gauge theory  is very complicated. They enter in the
dual operators as integers in the phase factors (\ref{phase}) and the
dependence of the two-point functions on these integers is determined
by sums of the type (\ref{sumJ}). The associated combinatorics is
extremely involved even for the simplest operators in this class
containing only one fermion bilinear. We shall therefore leave this
part of the  result (\ref{masscorr}) as a string theory prediction for
the instanton contribution to the dual two-point functions in the
gauge theory.

\subsection{Mixed \RR\!\!--\NSNS amplitudes}
\label{RRNSNSstring}

The instanton contributions to mixed \RR\!\!--\NSNS two-point
functions of section \ref{RRNSNS2pt} are dual to amplitudes of the
form
\be
\scrA^{k,h;l}_{\bfel,\bfn,\bfm;\bfel^\pp,\bfn^\pp}
= \gs^{7/2}\er^{2\pi i\tau}\, {}_{_1}\!\la\chi^{k,h}_{\bfel,\bfn,\bfm}|
\otimes{}_{_2}\!\la\chi^l_{\bfel^\pp,\bfn^\pp}
\|V_2\ra\!\ra \, ,
\label{mixstrampl}
\ee
where as external states we take
\ba
{}_{_1}\!\la\chi^{k,h}_{\bfel,\bfn,\bfm}|\otimes
{}_{_2}\!\la\chi^l_{\bfel^\pp,\bfn^\pp}|
&=& \veps_{i_1i_2i_3i_4}\,\veps_{i^\pp_1i^\pp_2i^\pp_3i^\pp_4}
\,\fr{\omega_{\ell_1}\omega_{\ell_2}\omega_{\ell_1^\pp}
\omega_{\ell_2^\pp}\omega_{n^\pp_1}\cdots\omega_{n^\pp_l}}
\label{extstatemix} \\
&\times& {}_h\!\la0|\a^{(1)i_1}_{\ell_1}\a^{(1)i_2}_{\ell_2}
\tilde\a^{(1)i_3}_{\ell_1}\tilde\a^{(1)i_4}_{\ell_2} \prod_{r=1}^k
\left[S_{n_r}^{(1)-}\tilde S_{n_r}^{(1)-} \right] \prod_{s=1}^h
\left[S_{m_s}^{(1)+}\tilde S_{m_s}^{(1)+} \right] \nn \\
&\otimes& {}_h\!\la0|\a^{(2)i^\pp_1}_{\ell_1^\pp}
\a^{(2)i^\pp_2}_{\ell_2^\pp}\tilde\a^{(2)i^\pp_3}_{\ell_1^\pp}
\tilde\a^{(2)i^\pp_4}_{\ell_2^\pp}
\prod_{u=1}^{l} \left[\a^{(2)j_u}_{n^\pp_u}
\tilde\a^{(2)j_u}_{n^\pp_u} \right] \, . \nn
\ea
The calculation of the amplitude (\ref{mixstrampl}) is very similar to
that of the previous subsection. One should distribute four broken
dynamical supersymmetries on each disk. The contribution of the first
disk is then exactly as in the previous \RR case. On the second disk
one should lower from the exponent two $S_{-r}M_r\tilde S_{-r}$
bilinears which after going through the supercharges annihilate the
two $\a$'s and the two $\tilde\a$'s in the external state which are
contracted via the $\veps$ tensor. Hence the contribution of these
oscillators to the amplitude is again the same as in the previous
case. The remaining pairs of bosonic oscillators in the external state
require that $l$ factors of $\fr{\omega_r}\a^j_{-r}\tilde\a^j_{-r}$
be lowered from the exponent in $\|V_2\ra\!\ra$. In the large $m$
limit
\be
\omega_r \sim m \, , \qquad [\a_r,\a_{-r}] \sim m \, , \qquad
[\tilde\a_r,\tilde\a_{-r}] \sim m \, , \qquad \forall r \, ,
\label{largemcomm}
\ee
so that the contribution of the $l$ remaining  pairs of bosonic
oscillators simply cancels $l$ factors of $m$ in the normalisation in
(\ref{extstatemix}). Notice that the only non-zero contribution is the
one just described. In particular, it is not possible to use the two
$S_{-r}M_r\tilde S_{-r}$ bilinears to annihilate pairs of external
oscillators with contracted SO(4)$_R$ indices. In this case
$\fr{\omega_r}\a^j_{-r}\tilde\a^j_{-r}$ factors  lowered from the
exponent would have to be commuted with the $\a$'s and $\tilde\a$'s
contracted into the $\veps$, but these commutators vanish for symmetry
reasons.

In conclusion the result for the amplitude (\ref{mixstrampl}) in the
large $m$ limit is
\be
\scrA^{k,h;l}_{\bfel,\bfn,\bfm;\bfel^\pp,\bfn^\pp}
\sim \gs^{7/2}\,\er^{2\pi i\tau} \,
m^{8+(k-h)} \,\fr{\ell_1\ell_2\ell_1^\pp\ell_2^\pp}\,
\frac{\prod_{j=1}^h m_j}{\prod_{i=1}^k n_i} \, .
\label{mixstrresult}
\ee
The rescaled contribution to the mass matrix is thus
\be
\fr{\mu} \,\d M \sim g_2^{7/2}\,
\er^{-\frac{8\pi^2}{g_2\lambda^\pp}+i\theta} \,
\fr{(\lambda^\pp)^{\half(k-h)}} \,
\fr{\ell_1\ell_2\ell_1^\pp\ell_2^\pp}\,
\frac{\prod_{j=1}^h m_j}{\prod_{i=1}^k n_i}
\label{mixmasscorr}
\ee
in agreement with the Yang--Mills result (\ref{mixgNJdepfin}). As in
the \RR example of the previous subsection, we also find a very simple
dependence on the mode numbers of the external states. Notably, the
result only depends on the mode numbers, $\ell_1^\pp$ and $\ell_2^\pp$,
of the four oscillators contracted via the $\veps$ tensor in the \NSNS
state and it is independent of the mode numbers, $n_u^\pp$,
$u=1,\ldots,l$, of the remaining oscillators.

From the calculation of the amplitude (\ref{mixstrampl}) we can
immediately deduce the result for amplitudes of the form
\ba
&& \hsp{-0.3}\scrA^{l}_{\bfel,\bfn;\bfel^\pp,\bfn^\pp}
= \gs^{7/2}\er^{2\pi i\tau}\,
\veps_{i_1i_2i_3i_4}\,\veps_{i^\pp_1i^\pp_2i^\pp_3i^\pp_4}
\,\fr{\omega_{\ell_1}\omega_{\ell_2}\omega_{\ell_1^\pp}
\omega_{\ell_2^\pp}\omega_{n_1}\cdots\omega_{n_l}
\omega_{n^\pp_1}\cdots\omega_{n^\pp_l}}
\label{extstatens} \\
&& \hsp{-0.3}{}_h\!\la0|\a^{(1)i_1}_{\ell_1}\a^{(1)i_2}_{\ell_2}
\tilde\a^{(1)i_3}_{\ell_1}\tilde\a^{(1)i_4}_{\ell_2}
\prod_{r=1}^l \!\left[\a^{(1)j_r}_{n_r}
\tilde\a^{(1)j_r}_{n_r} \right]
\!\otimes\!{}_h\!\la0|\a^{(2)i^\pp_1}_{\ell_1^\pp}
\a^{(2)i^\pp_2}_{\ell_2^\pp}\tilde\a^{(2)i^\pp_3}_{\ell_1^\pp}
\tilde\a^{(2)i^\pp_4}_{\ell_2^\pp}
\prod_{s=1}^{l} \!\left[\a^{(2)j^\pp_s}_{n^\pp_s}
\tilde\a^{(2)j^\pp_s}_{n^\pp_s} \right] \! \|V_2\ra\!\ra . \nn
\ea
which correspond to the two point functions with scalar impurities
(\ref{gennsns2ptf}) briefly discussed at the end of section
\ref{RRNSNS2pt}. Both disks in this case are treated as the second
disk in the calculation of the mixed \RR\!\!--\NSNS amplitude
(\ref{mixstrampl}). The only non trivial dependence on $m$ and on the
mode numbers comes from the eight oscillators contracted via the two
$\veps$ tensors, all the other oscillators and the associated
normalisation factors are simply cancelled by terms in the expansion
of the boundary state. The resulting contribution to the string mass
matrix is
\be
\fr{\mu} \,\d M \sim g_2^{7/2}\,
\er^{-\frac{8\pi^2}{g_2\lambda^\pp}+i\theta} \,
\fr{\ell_1\ell_2\ell_1^\pp\ell_2^\pp} \, ,
\label{nsmasscorr}
\ee
for any number, $l$, of oscillators in the external states. The
dependence on the parameters is again in agreement with the gauge
theory result (\ref{genscaladim}). The mode number dependence in
(\ref{nsmasscorr}) is very surprising from the Yang--Mills point of
view. The fact that the mass corrections, and thus the corresponding
anomalous dimensions, only depend on the first four mode numbers in
each state requires dramatic cancellations in the dual gauge theory
calculation and it would be interesting to verify this explicitly at
least for the simplest operators in this class corresponding to $l=1$.

\section{Perturbative mixing of the \NSNS and \RR sectors}
\label{perturb}

In the previous sections we discussed two-point correlation
functions in $\scrN$=4 SYM, as well as the corresponding plane-wave
string amplitudes, which vanish at tree level but receive non-zero
($D$-)instanton contributions. We will now see whether the
same processes might also receive perturbative contributions. In this
section we present a qualitative analysis of perturbative corrections
to \NSNS\!--\RR mixing processes of the type discussed in sections
\ref{RRNSNS2pt} and \ref{RRNSNSstring}. A similar analysis can be
repeated for the correlation functions and string amplitudes of
sections \ref{RR2pt} and \ref{RRstring}.

We first consider string loop corrections to a two-point amplitude
mixing \NSNS and \RR states, focusing on the simplest process of the
type (\ref{mixstrampl}), in which the incoming and outgoing states are
SO(4)$_C\times$SO(4)$_R$ singlets containing respectively two massive
fermionic oscillators and two massive bosonic oscillators. The
analysis of the one-loop string amplitude provides non-trivial
predictions for the dual Yang--Mills two-point function which will be
addressed in the following subsection.

\subsection{String perturbation theory}
\label{stringpert}

As an example of a string amplitude with mixing of the \NSNS and \RR
sectors we consider a two-point function coupling two impurity
states. Since we do not have to worry about fermionic zero modes as in
the ($D$-)instanton induced amplitudes, there is no need to include
additional bosonic oscillators in the external states. The states we
consider are  SO(4)$_C\times$SO(4)$_R$ singlets in the \RR sector,
\be
|\chi_n\ra_{_{\!1}}\!\!{}^{\scriptscriptstyle(\RRm)} = \veps_{ab}\,
\big(S^-_{-n}\big)^{\a a} \big(\tilde S^-_{-n}\big)^b_\a \, |0\ra_h\, ,
\label{2impstringRR}
\ee
and in the \NSNS sector,
\be
|\chi_n\ra_{_{\!2}}\!\!{}^{\scriptscriptstyle(\NSNSm)} = \fr{\omega_n}\,
\d_{ij} \, \a^i_{-n} \tilde\a^j_{-n} \, |0\ra_h \, .
\label{2impstringNSNS}
\ee
The quadratic string theory hamiltonian is diagonal in the bosonic and
fermionic oscillators so there is no tree level amplitude coupling the
states (\ref{2impstringRR}) and (\ref{2impstringNSNS}). We will argue,
however, that a non-zero two-point amplitude between these states can
arise at one loop in the plane-wave background, whereas it is absent
in flat space. We will only indicate the origin of this mixing since a
complete evaluation of the one-loop amplitude is beyond the scope of
this paper.

The one-loop string mass matrix between the two string states
(\ref{2impstringRR}) and (\ref{2impstringNSNS}) is given by
\begin{equation}
M_{12}= {}^{\scriptscriptstyle(\RRm)}\!\!{}_{_1}\!\la \chi_n |
\big[H_3 (E_n^{(0)}-H_2)^{-1} H_3+H_4 \big] |
\chi_n\ra_{_{\!2}}\!\!{}^{\scriptscriptstyle(\NSNSm)} \, ,
\label{1loopstrampl}
\end{equation}
where the first term represents gluing two cubic vertices with
propagators and summing over intermediate states, while the second
term represents a contact term whose form is dictated by
supersymmetry. The first term is schematically represented in figure
\ref{str-1loop-diagr}. The eigenvalues of the complete mass matrix in
this sector should be compared with the eigenvalues of the dilation
operator in the corresponding sector of the $\scrN$=4 Yang--Mills
theory.
\FIGURE[!h]{
\includegraphics[width=0.9\textwidth]{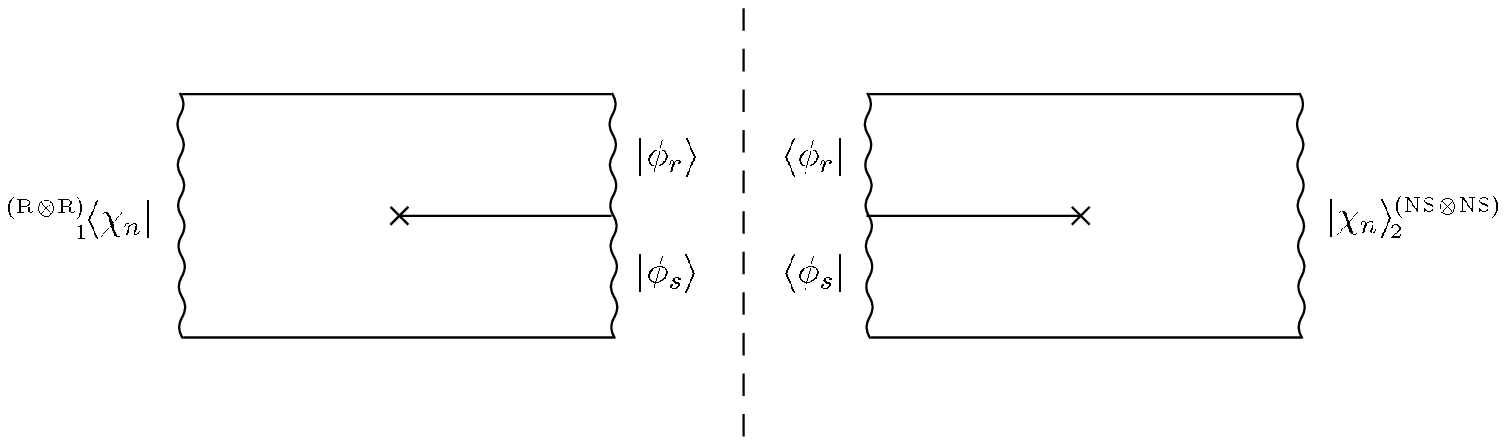}
\caption{String one-loop contribution.}
\label{str-1loop-diagr}
}
Figure \ref{str-1loop-diagr} indicates a sum over intermediate states
that couple to the external states via the cubic vertex.  In principle
this sum includes states with an arbitrary number of oscillators.  The
form of the plane-wave string cubic vertex \cite{pert} leads to
potentially non-zero contributions to (\ref{1loopstrampl}) in impurity
non-preserving channels. An example of such a contribution involves
the intermediate states
\ba
&& |\phi_r\ra \sim \big(\a \tilde S\big) |0\ra_h \nn \\
&& |\phi_s\ra \sim \big(\tilde\a S\big) |0\ra_h \, .
\label{interstates}
\ea
The structure of the string cubic vertex allows the coupling of these
states to the external states (\ref{2impstringRR}) and
(\ref{2impstringNSNS}). The process is permitted because the string
cubic hamiltonian in the plane-wave background does not factorise into
left- and right-moving parts. This is a feature which distinguishes
the string theory interactions in the plane-wave background from those
in flat space. In flat space, where mixing of \NSNS and \RR states
does not take place, the process just described is not possible
because of the factorisation of the interaction vertex.

The above example illustrates a mechanism which makes the perturbative
mixing of the \NSNS and \RR sectors possible in the plane-wave
background. The cancellation of all the contributions of the type
described here appears extremely unlikely, although a detailed one
loop analysis would be necessary to prove that matrix elements such as
(\ref{1loopstrampl}) are really non-zero.

\subsection{$\scrN$=4 SYM perturbation theory}
\label{sympert}

The arguments in the previous subsection strongly indicate that string
two-point amplitudes mixing the \NSNS and \RR sectors receive non-zero
perturbative contributions in the maximally supersymmetric plane-wave
background, unlike the corresponding processes in flat space. This
observation, combined with the vanishing of the same amplitudes at
tree level, has non-trivial implications for the two-point functions
of the dual operators in the BMN limit of the $\scrN$=4 Yang--Mills
theory. In the BMN correspondence the tree level result for a string
amplitude encompasses the whole planar perturbative expansion of the
gauge theory, \ie it is exact to all orders in the $\lambda^\pp$
expansion. String loop corrections correspond to non-planar
corrections in the gauge theory, with both sides being reorganised in
a series in powers of  $g_2$. Therefore the results of the previous
subsection predict that Yang--Mills correlation functions dual to
mixed \NSNS\!-\RR string amplitudes should be zero at all orders in
the planar approximation, but should receive non-zero perturbative
corrections at the non-planar level. In this section we show that this
is indeed the case, at the leading non-trivial order, for the
two-point function dual to the amplitude considered in the previous
subsection.

The operators dual to the string states (\ref{2impstringRR}) and
(\ref{2impstringNSNS}) are respectively of the form
\be
\scrO_1 = \frac{\veps_{ab}}{\sqrt{J\left(\frac{\gy^2N}{8\pi^2}
\right)^{J+2}}} \sum_{p=0}^J \er^{2\pi ipn/J}\,
\Tr\left(Z^{J-p}\psi^{-\,\a a}Z^p\psi^{- \,b}_\a \right)
\label{2impRRop}
\ee
and
\be
\bar\scrO_2 = \fr{\sqrt{J\left(\frac{\gy^2N}{8\pi^2}\right)^{J+3}}}
\sum_{q=0}^{J+1} \er^{-2\pi iqn/J} \, \Tr\left(\barZ^{J+1-q}\v^i
\barZ^q\v^i \right) \, .
\label{2impNSNSop}
\ee
The operator $\bar\scrO_2$ contains $J+1$ $\barZ$ fields so that it
has the same bare dimension as $\scrO_1$. This is reflected in the
power of $\gy^2N$ in the normalisation.  Notice that in this section
we are using the same conventions adopted in the rest of the paper,
which are not the standard ones used in perturbative calculations. In
our normalisations the Yang--Mills coupling appears in the action only
as an overall factor of $1/\gy^2$. Hence all the interaction vertices
are proportional to $1/\gy^2$ and all the propagators are proportional
to $\gy^2$.  With these conventions the normalisations of the
operators $\scrO_1$ and $\bar\scrO_2$ are such that the two-point
functions $\la\scrO_1\,\bar\scrO_1\ra$ and
$\la\scrO_2\,\bar\scrO_2\ra$  are of order 1 at tree level.

We are interested in perturbative corrections to the two-point
function
\be
G(x_1,x_2) = \la \scrO_1(x_1) \, \bar\scrO_2(x_2) \ra \, ,
\label{def2ptf}
\ee
which vanishes at tree level.

Let us first analyse the planar contributions. These correspond to
tree level amplitudes in string theory and thus are expected to
vanish. The leading perturbative contributions in the planar
approximation correspond to diagrams with the two distinct topologies
represented in figure \ref{topologies}.
\FIGURE[!h]{
\includegraphics[width=0.75\textwidth]{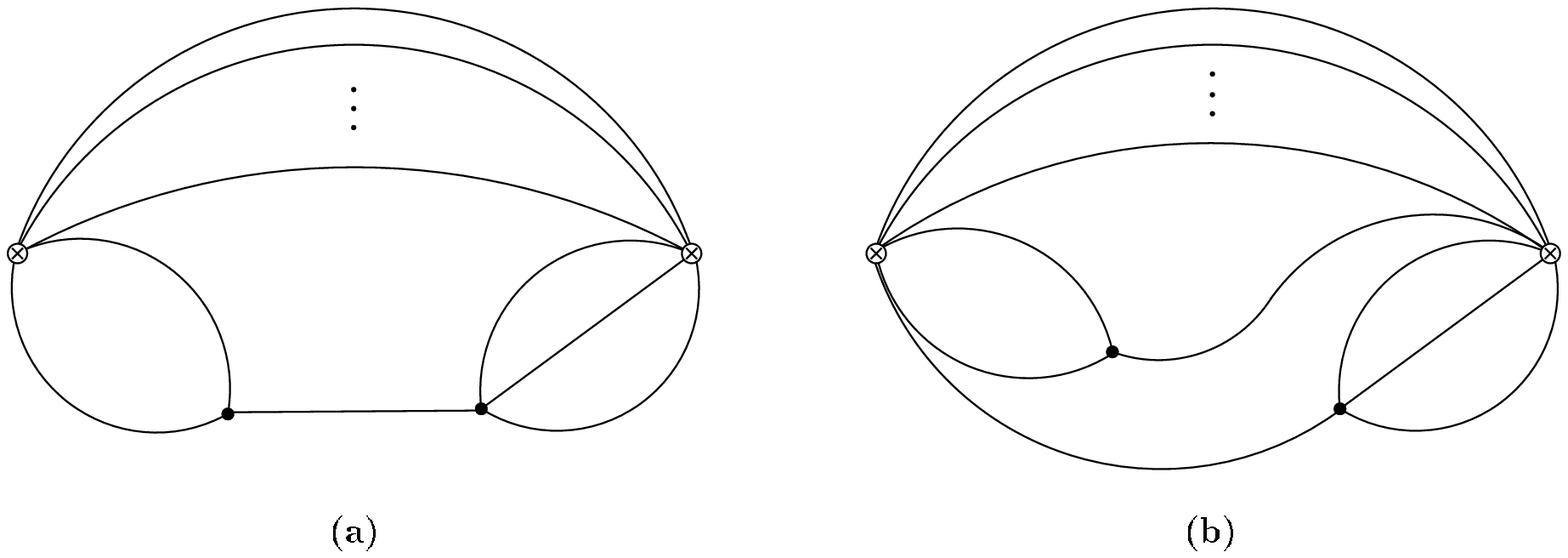}
\caption{Topologies of leading planar contributions.}
\label{topologies}
}

The couplings in the $\scrN$=4 lagrangian which are relevant for these
diagrams are
\be
\scrL_{\rm int} = \fr{\gy^2} \, \Tr\left(Z\left[
\bar\psi^{-\,2}_\adot,\bar\psi^{-\,\adot 3}\right] +
\left[Z,\v^i\right]\left[\barZ,\v^i\right] \right) \, .
\label{interaction}
\ee

We shall not compute explicitly the diagrams in figure
\ref{topologies}. The sum of the two types of contributions is
logarithmically divergent. For simplicity, in the following we shall
only discuss the combinatorics associated with diagrams of the
topology (a) in figure \ref{topologies}. Our considerations apply to
the diagrams of type (b) as well and it is understood that the two
types of contributions are included in the calculation of the
two-point function.

The planar diagrams in figure \ref{topologies} require $p=0$ in the
operator $\scrO_1$ (\ie no $Z$ lines can be present between the two
fermions) and $q=0$ or $q=1$ in the operator $\bar\scrO_2$ (there can
be at most one $\barZ$ between the two scalars in the
trace). Indicating with dashed lines the $\la Z\barZ\ra$ propagators,
with dotted lines the $\la\v\v\ra$ propagators and with plain lines
the fermion propagators, the relevant diagrams are those in figure
\ref{planar-diagr-a}. The first two diagrams involve the $q=0$ term in
the operator $\bar\scrO_2$, whereas the third diagram involves the
$q=1$ term.
\FIGURE[!h]{
\includegraphics[width=0.85\textwidth]{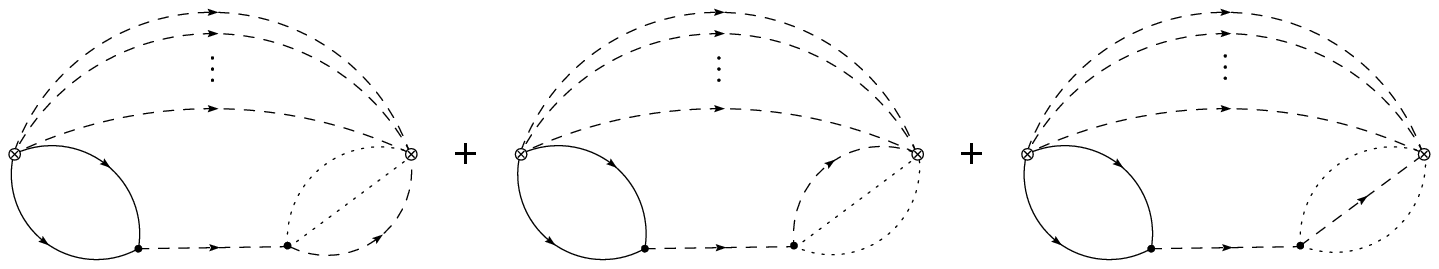}
\caption{Planar diagrams of type (a).}
\label{planar-diagr-a}
}

Taking into account the normalisation of the operators the sum of the
diagrams in figure \ref{planar-diagr-a} and the analogous ones
obtained from (b) in figure \ref{topologies} gives
\ba
&& \fr{\sqrt{J\left(\frac{\gy^2N}{8\pi^2}\right)^{J+2}}} \,
\fr{\sqrt{J\left(\frac{\gy^2N}{8\pi^2}\right)^{J+3}}} \,
\fr{\gy^4} \,\gy^{2(J+6)}\, N^{J+4}
\left(1-\er^{-2\pi i n/J} \right) f(x_1,x_2) \nn \\
&& \sim \frac{(2\pi i n) \,\gy^3 N^{3/2}}{J^2} \, f(x_1,x_2)
= \fr{J^{1/2}} \, (2\pi i n) \, \left(\lambda^\pp\right)^{3/2}
\, f(x_1,x_2) \, ,
\label{zero-planar}
\ea
where the logarithmically divergent function $f(x_1,x_2)$ is
determined integrating over the position of the interaction points.
In (\ref{zero-planar}) the power of $\gy$ results from the combination
of two interaction vertices, $J+6$ propagators and the normalisation
of the operators. The power of $N$ in the numerator in the first line
comes from the colour contractions. The factor
\be
\left(1-\er^{-2\pi in/J}\right)
\label{1-exp}
\ee
comes from the sum of the three diagrams in figure
\ref{planar-diagr-a}. The first two diagrams give the 1 (no
exponential because they correspond to $q=0$ in $\bar\scrO_2$) and the
third diagram gives the exponential term. It has weight 2 and a
relative minus sign with respect to the first two diagrams. Expanding
(\ref{1-exp}) for large $J$ gives the result in (\ref{zero-planar}),
which vanishes in the BMN limit. Therefore the leading planar
perturbative contributions vanish as expected.

Let us now consider the leading non-planar corrections to the
two-point function (\ref{def2ptf}). These correspond to string loop
corrections to the dual amplitude which are expected to be non-zero in
the plane-wave background. The leading non-planar corrections in the
gauge theory are suppressed by a factor of $1/N^2$ with respect to the
planar contributions. In order for the non-planar corrections to
survive in the BMN limit additional powers of $J$ should arise. There
are two sources of powers of $J$ in Feynman diagrams: the sums in the
definitions of the operators and the number of diagrams at each
genus. The operators (\ref{2impRRop})-(\ref{2impNSNSop}) involve one
sum each, so that potentially the sums can give a factor of
$J^2$. This, however, requires that the sums be independent and the
exponential factors in the operators be cancelled. It is easy to
verify that this is never the case. For operators containing $J$
elementary fields the number of diagrams at genus $g$ grows as
$J^{2g}$, so that again at the level of the leading non-planar
corrections one can potentially get a factor of $J^2$ adding diagrams
which give an equal contribution. This is what happens in the case of
the two-point function (\ref{def2ptf}). The relevant set of non-planar
diagrams is depicted in figure \ref{fig-nonplanar-nonzero}. A similar
set of diagrams is obtained from (b) in figure \ref{topologies}. The
number of diagrams in these series grows as $J^2$.
\FIGURE[!h]{
\includegraphics[width=0.85\textwidth]{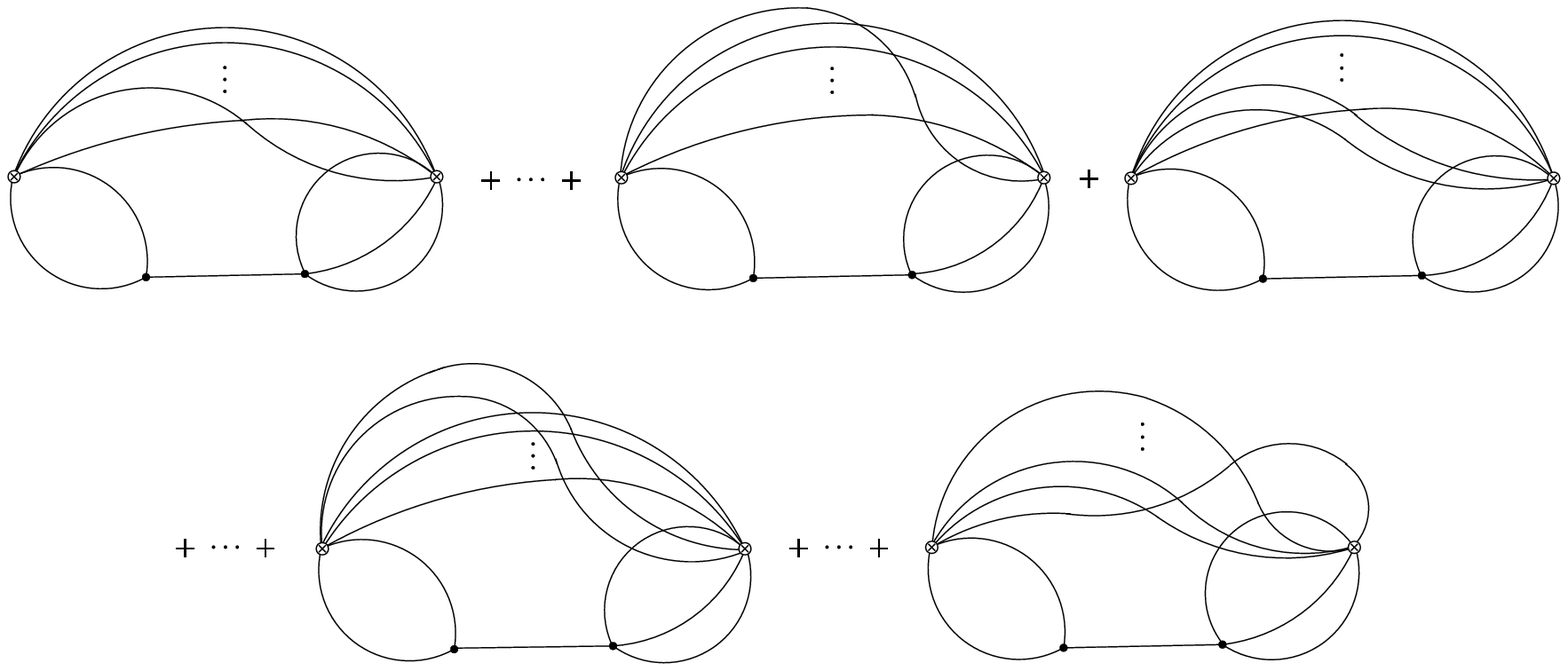}
\caption{Non-planar contributions surviving in the BMN limit.}
\label{fig-nonplanar-nonzero}
}

There are three sets of diagrams with the topologies in figure
\ref{fig-nonplanar-nonzero}. The three sets can be obtained as
non-planar deformations of the three diagrams in figure
\ref{planar-diagr-a}. Corresponding diagrams in the three series
differ in the number of $\barZ$ lines between the two $\v^i$
impurities in the operator $\bar\scrO_2$, \ie they involve different
terms in the sum in (\ref{2impNSNSop}). This implies that adding up
the three sets does not generate a factor such as (\ref{1-exp}) which
would give a $1/J$ suppression as in (\ref{zero-planar}).

In the case of the series obtained deforming the second diagram in
figure \ref{planar-diagr-a} all the diagrams correspond to $q=0$ in
the operator $\bar\scrO_2$, whereas in the other two series the
diagrams have $\barZ$ lines originating between the two $\v^i$ lines
and thus correspond to different values of $q$. The leading large-$N$
contribution from the sum of the three series corresponding to the
diagrams in figure \ref{fig-nonplanar-nonzero} and the analogous ones
obtained from (b) in figure \ref{topologies} is
\ba
&& \fr{\sqrt{J\left(\frac{\gy^2N}{8\pi^2}\right)^{J+2}}} \,
\fr{\sqrt{J\left(\frac{\gy^2N}{8\pi^2}\right)^{J+3}}} \, \fr{\gy^4}
\gy^{2(J+6)} N^{J+2} \nn \\
&& \times \sum_{k=0}^{J} \,(J-k)\left(1-2\,\er^{-2\pi i(k+1)n/J}
+\er^{-2\pi ikn/J}\right) f(x_1,x_2) \nn \\
&& \sim \frac{\gy^3\,J}{N^{1/2}} \,\left(\half+\frac{i}{2n\pi}\right)
\, f(x_1,x_2) = \left(\half+\frac{i}{2n\pi}\right)
\left(g_2\right)^2 \left(\lambda^\pp\right)^{3/2} \, f(x_1,x_2) \, ,
\label{nonzero-nonplanar}
\ea
where we have used
\be
\sum_{k=0}^{J} \,(J-k) \left(1-2\,\er^{-2\pi i(k+1)n/J}
+\er^{-2\pi ikn/J}\right)
\sim J^2\left(\half+\frac{i}{2n\pi}\right)
\label{simplisum}
\ee
in the large $J$ limit.

Therefore the two-point function (\ref{def2ptf}) receives a
non-vanishing contribution at the leading non-planar level in the BMN
limit. The induced contribution to the matrix of anomalous dimensions
is of order $(g_2)^2\,(\lambda^\pp)^{3/2}$.

Elements of the matrix of anomalous dimensions corresponding to
non-real operators, such as those that we have considered, are in
general complex. This is the case for the contribution extracted from
the coefficient in (\ref{nonzero-nonplanar}) as well as for the
vanishing planar contribution (\ref{zero-planar}). The matrix element
corresponding to the conjugate operators is the complex conjugate of
the one computed here, so that the resulting matrix is hermitian and
has real eigenvalues corresponding to the physical scaling dimensions
of the operators. Notice also that, although half-integer powers of
$\lambda^\pp$ appear in two-point functions mixing operators with
fermionic and bosonic impurities, the anomalous dimensions obtained
resolving the mixing have an expansion in integer powers of
$\lambda^\pp$.

In this section we have presented a qualitative analysis of the
leading perturbative contributions to a two-point function with mixing
of the \NSNS and \RR sectors. Similar considerations can be repeated
for the string amplitudes of the type described in section
\ref{RRstring} and the dual gauge theory correlation functions of
section \ref{RR2pt}. String amplitudes of the form
(\ref{extstate})-(\ref{rrstrampl}) with $k\ne k^\pp$ and $h\ne h^\pp$
vanish at tree level, but are expected to receive a non-zero
contribution at one loop. Therefore the dual two-point functions
(\ref{gen2ptf}) should have the same behaviour as the mixed ones, \ie
they should vanish in the planar approximation at all orders in
$\lambda^\pp$, but they should receive non-vanishing corrections
beyond the zeroth order in $g_2$.

\acknowledgments{AS acknowledges financial support from PPARC and
Gonville and Caius college, Cambridge. We also wish to acknowledge
support from the European Union Marie Curie Superstrings Network
MRTN-CT-2004-512194.}


\begin{thebibliography}{123}


\bibitem{penrose}{M.~Blau, J.~Figueroa-O'Farrill, C.~Hull and
G.~Papadopoulos, ``A new maximally supersymmetric background of IIB
superstring theory'', \jhep{01}{2002}{047} [\hepth{0110242}];
``Penrose limits and maximal supersymmetry'', \cqg{19}{2002}{L87}
[\hepth{0201081}].}

\bibitem{bmn}{D.~Berenstein, J.~M.~Maldacena and H.~Nastase,
``Strings in flat space and pp waves from $\scrN$=4 super Yang Mills'',
\jhep{04}{2002}{013} [\hepth{0202021}].}

\bibitem{gks1}{M.B.~Green, S.~Kovacs and A.~Sinha,
``Non-perturbative contributions to the plane-wave string mass
matrix'', \jhep{05}{2005}{055} [\hepth{0503077}].}

\bibitem{gks2}{M.B.~Green, S.~Kovacs and A.~Sinha, ``Non-perturbative
effects in the BMN limit of $\scrN$=4 supersymmetric Yang-Mills'',
\hepth{0506200}.}

\bibitem{bgkr}{M. Bianchi, M.B. Green, S. Kovacs and G.C. Rossi,
``Instantons in supersymmetric Yang-Mills and $D$-instantons in IIB
superstring theory'', \jhep{08}{1998}{013}, [\hepth{9807033}].}

\bibitem{dhkmv}{N.~Dorey, T.J.~Hollowood, V.V.~Khoze, M.P.~Mattis and
S.~Vandoren, ``Multi-instanton calculus and the AdS/CFT
correspondence in $\scrN$=4 superconformal field theory'',
\npb{552}{1999}{88} [\hepth{9901128}].}

\bibitem{gk}{M.B.~Green and S.~Kovacs, ``Instanton-induced Yang--Mills
correlation functions at large $N$ and their  AdS$_5 \times S^5$
duals'', \jhep{04}{2003}{058} [\hepth{0212332}].}

\bibitem{pert}{M.~Spradlin and A.~Volovich, ``Superstring interactions
in a pp-wave background'', \prd{66}{2002}{086004} [\hepth{0204146}]. \\
M.~Spradlin and A.~Volovich, ``Superstring interactions in a pp-wave
background. II'', \jhep{01}{2003}{036} [\hepth{0206073}]. \\
A.~Pankiewicz and B.~Stefanski, ``pp-wave light-cone superstring field
theory'', \npb{657}{2003}{79} [\hepth{0210246}]. \\
A.~Pankiewicz, ``More comments on superstring interactions in the
pp-wave background'', \jhep{09}{2002}{056} [\hepth{0208209}]. \\
R.~Roiban, M.~Spradlin and A.~Volovich,
``On light-cone SFT contact terms in a plane wave'',
\jhep{10}{2003}{055} [\hepth{0211220}]. \\
J.~H.~Schwarz, ``The three-string vertex for a plane-wave background'',
[\hepth{0312283}]. \\
J.~Lucietti, S.~Schafer-Nameki and A.~Sinha, ``On the plane-wave cubic
vertex'', \prd{70}{2004}{026005} [arXiv:hep-th/0402185]; ``On the
exact open-closed vertex in plane-wave light-cone string field
theory'', \prd{69}{2004}{086005} [\hepth{0311231}]. \\
M.~Petrini, R.~Russo and A.~Tanzini,
``The 3-string vertex and the AdS/CFT duality in the pp-wave limit'',
\cqg{21}{2004}{2221} [\hepth{0304025}]. \\
G.~Grignani, M.~Orselli, B.~Ramadanovic, G.~W.~Semenoff and D.~Young,
``Divergence cancellation and loop corrections in string field theory
on a plane wave background'', \hepth{0508126}.}

\bibitem{recentgauge}{G.~Georgiou and G.~Travaglini, ``Fermion BMN
operators, the dilatation operator of $\scrN$=4 SYM, and pp-wave
string interactions'', \jhep{04}{2001}{001} [\hepth{0403188}]. \\
G.~Georgiou, V.~V.~Khoze and G.~Travaglini, ``New tests of the
pp-wave correspondence'', \jhep{10}{2003}{049} [\hepth{0306234}]. \\
P.~Matlock and K.~S.~Viswanathan, ``Four-impurity operators and string
field theory vertex in the BMN correspondence'', \prd{71}{2005}{026001}
[Erratum-\ibid{71}{2005}{029902}] [\hepth{0406061}]. \\
N.~Beisert, C.~Kristjansen, J.~Plefka and M.~Staudacher, ``BMN gauge
theory as a quantum mechanical system'', \plb{558}{2003}{229}
[\hepth{0212269}]. \\
J.~Pearson, M.~Spradlin, D.~Vaman, H.~Verlinde and A.~Volovich,
``Tracing the string: BMN correspondence at finite $J^2/N$'',
\jhep{05}{2003}{022} [\hepth{0210102}].
}

\bibitem{kpss}{C.~Kristjansen, J.~Plefka, G.~W.~Semenoff and
M.~Staudacher, ``A new double-scaling limit of $\scrN$=4 super
Yang-Mills theory and PP-wave strings'', \npb{643}{2002}{3}
[\hepth{0205033}].}

\bibitem{mit1}{N.~R.~Constable, D.~Z.~Freedman, M.~Headrick,
S.~Minwalla, L.~Motl, A.~Postnikov and W.~Skiba,
``PP-wave string interactions from perturbative Yang-Mills theory'',
\jhep{07}{2002}{017} [arXiv:hep-th/0205089].}

\bibitem{bkpss}{N.~Beisert, C.~Kristjansen, J.~Plefka, G.W.~Semenoff
and M.~Staudacher, ``BMN correlators and operator mixing in $\scrN$=4
super Yang-Mills theory'', \npb{650}{2003}{125} [\hepth{0208178}].}

\bibitem{mit2}{N.~R.~Constable, D.~Z.~Freedman, M.~Headrick and
S.~Minwalla, ``Operator mixing and the BMN correspondence'',
\jhep{10}{2002}{068} [\hepth{0209002}].}

\bibitem{sz}{A.~Santambrogio and D.~Zanon, ``Exact anomalous
dimensions of $\scrN$=4 Yang-Mills operators with large R charge'',
\plb{545}{2002}{425} [\hepth{0206079}].}

\bibitem{gks3}{M.~B.~Green, S.~Kovacs and A.~Sinha,
``Non-perturbative contributions in the plane-wave/BMN limit'',
\hepth{0510166}}.

\bibitem{bers}{M.~Bianchi, B.~Eden, G.~Rossi and Y.~S.~Stanev,
``On operator mixing in $\scrN$=4 SYM'',
\npb{646}{2002}{69} [\hepth{0205321}].}

\bibitem{met}{R.~R.~Metsaev, ``Type IIB Green-Schwarz superstring in
plane wave Ramond-Ramond background,'' \npb{625}{2002}{70}
[\hepth{0112044}].}

\bibitem{mt}{R.~R.~Metsaev and A.~A.~Tseytlin,
``Exactly solvable model of superstring in plane wave Ramond-Ramond
background'', \prd{65}{2002}{126004} [\hepth{0202109}].}

\bibitem{bei}{N.~Beisert, ``BMN operators and superconformal
symmetry'', \npb{659}{2003}{79} [\hepth{0211032}].}

\bibitem{sk}{S.~Kovacs, ``On instanton contributions to anomalous
dimensions in $\scrN$=4  supersymmetric Yang-Mills theory'',
\npb{684}{2004}{3} [\hepth{0310193}].}

\bibitem{akmrv}{D.~Amati, K.~Konishi, Y.~Meurice, G.C.~Rossi and
G.~Veneziano, ``Nonperturbative aspects in supersymmetric gauge
theories'', \prep{162}{1988}{169}.}

\bibitem{dhkm}{N.~Dorey, T.J.~Hollowood, V.V.~Khoze and
M.P.~Mattis, ``The calculus of many instantons'',
\prep{371}{2002}{231} [\hepth{0206063}].}

\bibitem{bvv}{A.V.~Belitsky, S.~Vandoren and P.~van Nieuwenhuizen,
``Yang-Mills and $D$-instantons'', \cqg{17}{2000}{3521}
[\hepth{0004186}].}

\bibitem{gg}{M.~R.~Gaberdiel and M.~B.~Green, ``The $D$-instanton and
other supersymmetric $D$-branes in IIB plane-wave string theory'',
{\it Annals Phys.} {\bf 307} (2003) 147 [\hepth{0211122}].}

\end{thebibliography}
\end{document}